\documentclass[sigconf,nonacm]{acmart}
\usepackage{amsmath}
\usepackage{amsfonts}
\usepackage{amstext}
\usepackage{graphicx}
\usepackage{textcomp}
\usepackage{xcolor}
\usepackage{url}           
\usepackage{hyperref}
\usepackage{booktabs}      
\usepackage{mathrsfs}
\usepackage{graphics}
\usepackage{balance}
\usepackage{caption}
\usepackage{multirow}
\usepackage{boldline}
\usepackage{wrapfig}
\usepackage{enumitem}
\usepackage{color,colortbl}
\usepackage{soul}
\usepackage{makecell}
\usepackage{hhline}
\usepackage{subfigure}
\usepackage{wrapfig}
\usepackage{algorithm}
\usepackage[noend]{algpseudocode}
\usepackage{eqparbox}
\usepackage{lipsum}
\usepackage{xargs}   
\usepackage[normalem]{ulem}

\AtBeginDocument{%
  \providecommand\BibTeX{{%
    \normalfont B\kern-0.5em{\scshape i\kern-0.25em b}\kern-0.8em\TeX}}}

\newcommand\modelName{DILI}
\newcommand\ModelWOLO{DILI-LO}
\newcommand\ModelWOAD{DILI-AD}
\newtheorem{definition}{\bf Definition}
\newcommand{\tabincell}[2]{\begin{tabular}{@{}#1@{}}#2\end{tabular}}
\DeclareMathOperator*{\argmin}{argmin}

\newdimen{\algindent}
\algnewcommand\LeftComment[2]{%
	\hspace{#1\algindent}$\triangleright$ \eqparbox{COMMENT}{#2} \hfill %
}
\newcolumntype{C}[1]{>{\centering\arraybackslash}p{#1}}
\newcolumntype{L}[1]{>{\raggedright}p{#1}}

\definecolor{red1}{RGB}{245,222,179}
\definecolor{red2}{RGB}{237,145,33}
\definecolor{red3}{RGB}{255,128,0}
\definecolor{grey}{RGB}{180,180,180}
\definecolor{green1}{HSB}{80,100,360}
\definecolor{green2}{HSB}{80,115,360}
\definecolor{green3}{HSB}{80,135,360}
\definecolor{green4}{HSB}{80,360,360}

\def\includeAppendix{1}

\newcommand\blfootnote[1]{%
	\begingroup
	\renewcommand\thefootnote{}\footnote{#1}%
	\addtocounter{footnote}{-1}%
	\endgroup
}

\if\includeAppendix0
\newcommand\vldbdoi{XX.XX/XXX.XX}
\newcommand\vldbpages{XXX-XXX}
\newcommand\vldbvolume{16}
\newcommand\vldbissue{9}
\newcommand\vldbyear{2023}
\newcommand\vldbauthors{\authors}
\newcommand\vldbtitle{\shorttitle} 
\newcommand\vldbavailabilityurl{https://github.com/pfl-cs/DILI}
\newcommand\vldbpagestyle{empty} 
\else
\fi

\begin{document}
 \if\includeAppendix0
 \title{\modelName: A Distribution-Driven Learned Index}
\else
\title{\modelName: A Distribution-Driven Learned Index (Extended version)}
\fi

\if\includeAppendix1
\author{Pengfei Li}
\affiliation{
		\institution{Alibaba Group, China}
		\country{}
	}
\email{lpf367135@alibaba-inc.com}

\author{Hua Lu}
\affiliation{
		\institution{Roskilde University, Denmark}
		\country{}
	}
\email{luhua@ruc.dk}

\author{Rong Zhu}
\affiliation{
		\institution{Alibaba Group, China}
		\country{}
	}
\email{red.zr@alibaba-inc.com}

\author{Bolin Ding}
\affiliation{
		\institution{Alibaba Group, China}
		\country{}
	}
\email{bolin.ding@alibaba-inc.com}

\author{Long Yang}
\affiliation{
		\institution{Peking University, China}
		\country{}
	}
\email{yanglong001@pku.edu.cn}

\author{Gang Pan}
\affiliation{
		\institution{Zhejiang University, China}
		\country{}
	}
\email{gpan@zju.edu.cn}

\else
\author{
	Pengfei Li{$^{1}$}, Hua Lu{$^{2\#}$}, Rong Zhu{$^{1}$}, Bolin Ding{$^{1}$}, Long Yang{$^{3}$} and Gang Pan{$^{4\#}$}}
\affiliation{%
	\institution{{$^{1}$}Alibaba Group, China, $^{2}$Roskilde University, Denmark, $^{3}$Peking University, China, $^{4}$Zhejiang University, China}
}
\affiliation{%
	\institution{{$^{1}$}\{lpf367135, red.zr, bolin.ding\}@alibaba-inc.com, $^{2}$luhua@ruc.dk, $^{3}$yanglong001@pku.edu.cn, $^{4}$gpan@zju.edu.cn}
}
\fi

\begin{abstract}
Targeting in-memory one-dimensional search keys, we propose a novel DIstribution-driven Learned Index tree (\textbf{\modelName}), where a concise and computation-efficient linear regression model is used for each node. 
An internal node's key range is equally divided by its child nodes such that a key search enjoys perfect model prediction accuracy to find the relevant leaf node. 
A leaf node uses machine learning models to generate searchable data layout and thus accurately predicts the data record position for a key.
To construct~\modelName, we first build a bottom-up tree with linear regression models according to global and local key distributions. Using the bottom-up tree, we build~\modelName~in a top-down manner, individualizing the fanouts for internal nodes according to local distributions.~\modelName~strikes a good balance between the number of leaf nodes and the height of the tree, two critical factors of key search time.
Moreover, we design flexible algorithms for~\modelName~to efficiently insert and delete keys and automatically adjust the tree structure when necessary.
Extensive experimental results show that~\modelName~outperforms the state-of-the-art alternatives on different kinds of workloads.
\if\includeAppendix1
\blfootnote{Hua Lu and Gang Pan are the corresponding authors.}
\fi
\end{abstract}

\maketitle

\if\includeAppendix0
\pagestyle{\vldbpagestyle}
\begingroup\small\noindent\raggedright\textbf{PVLDB Reference Format:}\\
\vldbauthors. \vldbtitle. PVLDB, \vldbvolume(\vldbissue): \vldbpages, \vldbyear.\\
\href{https://doi.org/\vldbdoi}{doi:\vldbdoi}
\endgroup
\begingroup
\renewcommand\thefootnote{}\footnote{\noindent
	$^{\#}$Corresponding authors

	\noindent\rule{.475\textwidth}{0.4pt}
	
	\noindent
	This work is licensed under the Creative Commons BY-NC-ND 4.0 International License. Visit \url{https://creativecommons.org/licenses/by-nc-nd/4.0/} to view a copy of this license. For any use beyond those covered by this license, obtain permission by emailing \href{mailto:info@vldb.org}{info@vldb.org}. Copyright is held by the owner/author(s). Publication rights licensed to the VLDB Endowment. \\
	\raggedright Proceedings of the VLDB Endowment, Vol. \vldbvolume, No. \vldbissue\ %
	ISSN 2150-8097. \\
	\href{https://doi.org/\vldbdoi}{doi:\vldbdoi} \\
}\addtocounter{footnote}{-1}\endgroup

\ifdefempty{\vldbavailabilityurl}{}{
	\vspace{.3cm}
	\begingroup\small\noindent\raggedright\textbf{PVLDB Artifact Availability:}\\
	The source code, data, and/or other artifacts have been made available at \url{\vldbavailabilityurl}.
	\endgroup
}
\fi

\section{Introduction}
\label{sec:intro}

\begin{figure*}[!htb]
	\centering
	\includegraphics[width=0.95\textwidth, trim={2mm 8mm 2mm 9mm},clip]{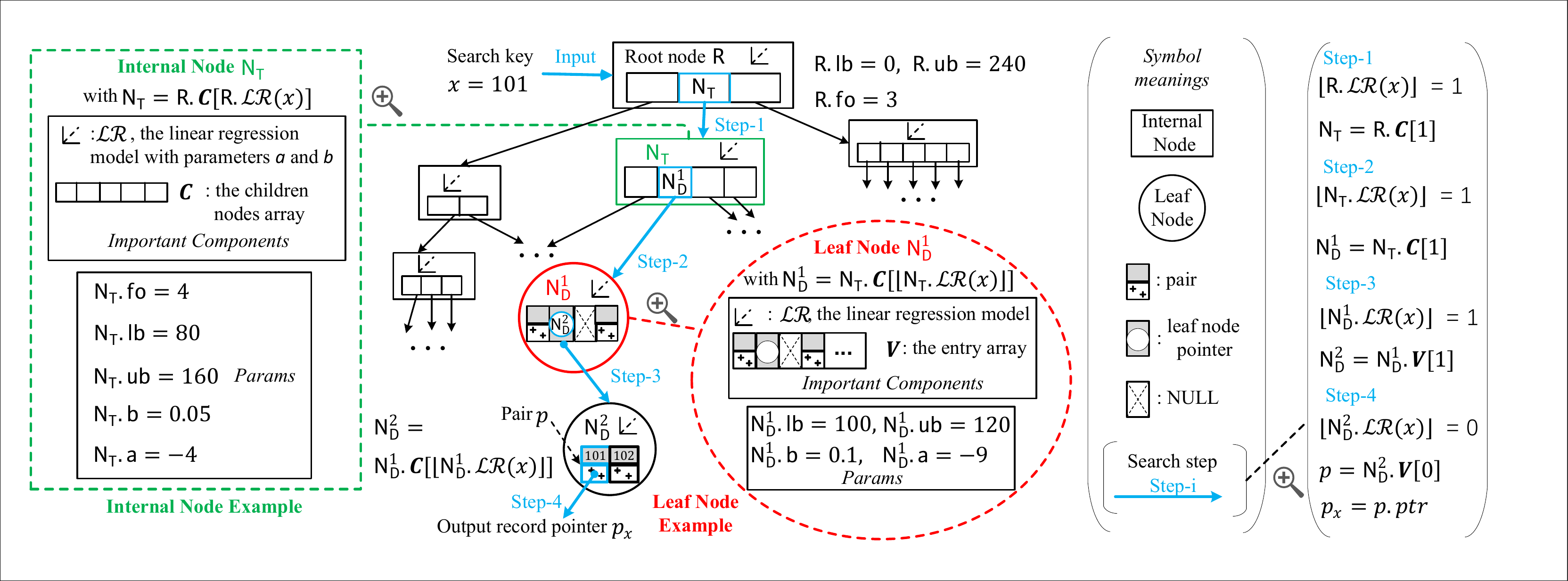}
	\caption{The Structure of~\modelName} 
	\label{fig:\modelName}
\end{figure*}

Recently, the learned index~\cite{DBLP:conf/sigmod/KraskaBCDP18} is proposed to replace B+Tree~\cite{DBLP:journals/csur/Comer79} in database search. It stages machine learning models into a hierarchy called Recursive Model Index (RMI). Given a search key $x$, RMI predicts, with some error bound, where $x$'s data is positioned in a memory-resident dense array.
Compared to B+Tree, RMI achieves comparable and even better search performance.
However, the layout of RMI, \emph{i.e.}, the number of stages and the number of models at each stage, must be fixed before the models are created.
Also, RMI fails to support key insertions and deletions.

To support data updates, ALEX~{\cite{DBLP:conf/sigmod/DingMYWDLZCGKLK20}} extends RMI by using a gapped array layout for the leaf level models. Moreover, ALEX uses cost models to initiate the RMI structure and to dynamically adapt the structure to updates. 
However, the stage layout of ALEX is not flexible enough as its fanout, \emph{i.e.}, the number of a node's child models, is stipulated to a power of 2. 
This renders ALEX's internal nodes' key ranges relatively static, which may result in node layout not good for particular key distributions, \emph{e.g.,} lognormal distribution. 
Also, ALEX's leaf level learned models do not guarantee accurate predictions. Thus, extra local search is needed to locate the required data, which downgrades the search performance.
More recently, LIPP~{\cite{DBLP:journals/pvldb/WuZCCWX21}} trains learned models for the whole dataset and places data at the predicted positions. When multiple data records are assigned to the same position, a new node is created at the position to hold them.
However, this simple strategy ignores the data distribution and often results in long traversal paths. Also, compared to B+Tree and ALEX, LIPP consumes much more memory.

In this paper, we design a novel index tree---DIstribution-driven Learned Index (\modelName). Its each node features an individualized fanout and a model created for a data portion whose key sequence is covered by the node's range.
For an internal node, its child nodes equally divide its range. Thus, the cost is minimized to locate the relevant leaf node in a key search. 
In a leaf node, an entry array $\boldsymbol{V}$ holds the keys in the node's range and the pointers to the corresponding data records. In addition, a leaf node uses an efficient linear regression model to map its keys to the positions in $\boldsymbol{V}$.

A critical issue for constructing~\modelName~is to determine its node layout that is able to achieve good search performance.
We design a sophisticated approach aware of data distributions and search costs.
A key search in~\modelName~involves two steps: 1) finding the leaf node covering the given key and 2) local search inside the leaf node.
Accordingly, the general search performance depends on two factors: leaf nodes' depths and linear regression models' accuracy in the leaf nodes.
Both factors should be considered in~\modelName~construction.

To this end, we propose a two-phase bulk loading approach. 
The first phase creates a distribution-driven bottom-up tree (BU-Tree),  whose node layout is determined by a greedy merging algorithm that considers both aforementioned factors. The merging creates linear regression models, starting at the bottom level to fully utilize the known key distribution. As a result, the models in the BU-Tree's leaf nodes guarantee good accuracy.
Basically, we build~\modelName~by making its node layout overall similar to that of the BU-Tree.
However, a BU internal node's range is not necessarily equally divided by its child nodes. Therefore, search in the BU-Tree's internal nodes can incur extra time to decide which child node to visit. To this end, the second phase converts the BU-Tree to a~\modelName~by redistributing keys among sibling nodes. When doing so, we carefully set different fanouts for~\modelName's different internal nodes according to their local key distributions, such that each internal node is equally divided by its child nodes.
Meanwhile, we retain good model accuracy in~\modelName's leaf nodes and keep them at the same level as the counterparts in the BU-Tree. 
As a result, we obtain a~\modelName~that is efficient at finding leaf nodes and have linear regression models of high accuracy in leaf nodes.
In other words, we first create a mirror model (BU-Tree) that exhibits a good node layout but cannot guarantee perfect accuracy, and then we create another similar model (\modelName) that avoids the mirror model's drawbacks but maintains its advantages.

It is noteworthy that we build the BU-Tree and~\modelName~according to detailed analyses of search costs, which consider caching effects in the main-memory context. Though~\modelName~and ALEX share some structural similarities, they are constructed according to different perspectives. ALEX is built top down, while the BU-Tree is bottom up and initially deals with all the keys. Thus, the BU-Tree understands the key distribution better and  partitions them into leaf nodes more reasonably. 
This make the `mirrored'~\modelName~achieve good local search performance. 
Also, our proposed cost function makes the BU-Tree (and~\modelName) have a suitable height. 
As a result, finding leaf nodes in~\modelName~consumes less time.

Our bulk loading approach makes the linear regression models in~\modelName's leaf nodes have high but not 100\% accuracy. 
We find that the `last-mile' local search in the leaf nodes is often the bottleneck of an entire query. To this end, we conduct a local optimization at each leaf node after the bulk loading, forcibly making the key-to-position mapping precise.
If multiple keys are mapped into the same position, a new child node is created to hold them. Experimental results show that local optimizations improve the query performance of~\modelName~by avoiding the local search inside the leaf nodes. 

Our local optimization is inspired by LIPP~\cite{DBLP:journals/pvldb/WuZCCWX21} and LISA~\cite{DBLP:conf/sigmod/Li0ZY020}. However, unlike LIPP,~\modelName's local optimization applies to leaf nodes only. Also, the two phase bulk loading approach makes~\modelName~reasonably partitions data such that the keys covered by leaf nodes are almost linearly distributed. Compared to LIPP, the linear regression models in~\modelName's leaf nodes assign fewer keys the same slots. Thus,~\modelName~encounters less conflicts and achieves better search performance and lower memory consumption.

Furthermore, ~\modelName~supports data updates. 
When an inserted key conflicts with an existing key at a data slot of~\modelName's leaf node, our insertion algorithm creates a new leaf node to cover the conflicting data.
Also,~\modelName~redistributes data covered by a leaf node in a balanced way, when insertions generate too many nodes and degrades the query performance.
Meanwhile, it allocates more data slots for those leaf nodes that encounter more frequent conflicts. In this way,~\modelName's height is bounded and the query performance downgrades slightly even for many insertions.
In addition, when a leaf node covers only one key after some deletions, this node will be trimmed to improve performance and save memory consumption.

We make the following major contributions in this paper.

\begin{itemize}[leftmargin=*]
	\item We design a distribution-driven learned index~\modelName~for in-memory 1D keys, together with algorithms and cost analysis.
	
	\item Accordingly, we design a distribution-driven BU-Tree as a node layout reference for~\modelName, and after more specific cost analyses we devise an algorithm to construct~\modelName~based on BU-Tree.
	
	\item We propose a local optimization on~\modelName's leaf nodes to avoid the local search and improve query performance.
	
	\item To update~\modelName~for key insertions and deletions, we devise efficient algorithms that retain search performance.
	
	\item We experimentally validate~\modelName's performance advantage over state-of-the-art alternatives on synthetic and real datasets.
\end{itemize}
	
The rest of the paper is organized as follows. Section~\ref{sec:structure} gives an overview of~\modelName. 
Section~\ref{sec:cost_analysis} analyses its search cost.
Sections~\ref{sec:bulk_load},~\ref{sec:optimization} and~\ref{sec:data_update} elaborate on~\modelName's construction, local optimization and updates, respectively.
Section~\ref{sec:experiments} reports on the experimental studies.
Section~\ref{sec:related_work} reviews the related work. Section~\ref{sec:conclusion} concludes the paper.


\section{Overview of~\modelName}
\label{sec:structure}

Table~\ref{tab:notations} lists the important notations used in the paper.
\begin{smaller}
	\begin{table}[htb]
		\if\includeAppendix0
		\smaller
		\fi
		\centering
		\caption{Notations}\label{tab:notations}
		\begin{tabular}{L{1.2cm}|l}
			\hline 
			\tabincell{l}{$\mathsf{N}.\mathtt{fo}$} & \tabincell{l}{Fanout of the node $\mathsf{N}$. $\mathsf{N}$ can be an internal or a leaf node.} \\ \hline
			\tabincell{l}{$\mathsf{N}.\mathcal{LR}$} & \tabincell{l}{Linear regression model of the node $\mathsf{N}$} \\ \hline
			\tabincell{l}{$\mathsf{N_T}.\boldsymbol{C}$} & \tabincell{l}{The child node array of the internal node $\mathsf{N_T}$} \\ \hline
			\tabincell{l}{$\mathsf{N_D}.\boldsymbol{V}$} & \tabincell{l}{The entry array of the leaf node $\mathsf{N_D}$} \\ \hline
			\if\includeAppendix1
			$\mathsf{N_D}.\Omega$ & Number of pairs covered by the leaf node $\mathsf{N_D}$ \\ \hline
			$\mathsf{N_D}.\Delta$ & \tabincell{l}{Total number of entries to be accessed to search for all keys \\ covered by $\mathsf{N_D}$, starting from $\mathsf{N_D}$} \\ \hline
			$\mathsf{N_D}.\kappa$ & \tabincell{l}{Average number of entries to be accessed to search for a\\key, starting from $\mathsf{N_D}$,  after the last local optimization to $\mathsf{N_D}$} \\ \hline
			$\mathsf{N_D}.\alpha$ & Number of adjustments of $\mathsf{N_D}$ so far \\ \hline
			\fi
			$\boldsymbol{\mathrm{T_{s}}}(x)$ & Search cost of key $x$ in~\modelName~without local optimization. \\ \hline
			$\boldsymbol{\mathrm{T_{ns}^{B}}}(\mathsf{N}, x, h)$ & Search cost of key $x$ \emph{w.r.t.}  a BU node $\mathsf{N}$ at height $h$ \\ \hline
			$\boldsymbol{\mathrm{T_{ea}^{B}}}(\boldsymbol{X}_{h}, \boldsymbol{X})$ & \tabincell{l}{Estimated accumulated search cost of the break points list \\ $\boldsymbol{X}_{h}$ for the key set $\boldsymbol{X}$ in the BU-Tree}  \\ \hline
		\end{tabular}
	\end{table}
\end{smaller}

\begin{definition} [Pair]
	\emph{A pair is a 2-tuple $p = (key, ptr)$, where $ptr$ is a pointer to the data record identified by $key$. }
\end{definition}

Let $\boldsymbol{P} = [p_{0}, p_{1}, \cdots, p_{|\boldsymbol{P}| - 1}]$  be an array of pairs, and \textproc{keys}($\boldsymbol{P}$) = $[p_{0}.key,p_{1}.key, \cdots, p_{|\boldsymbol{P}|-1}.key]$ the key sequence from $\boldsymbol{P}$.

\begin{definition} [Least square estimator]
	\emph{Given $I \subseteq [\tilde{n}] = \{0,1,$ $\cdots,n-1\}$, two sequences $\boldsymbol{X} = [x_{0},\cdots,x_{n-1}]$ and $\boldsymbol{Y} = [y_{0},\cdots,y_{n-1}]$, the least square estimator restricted to $I$ is the linear function that minimizes $\sum_{i \in I}(y_{i}-f(x_{i}))^{2}$ over any linear function $f$.  We use \textproc{leastSquares}($\boldsymbol{X}, \boldsymbol{Y}, I$) to denote an algorithm that finds the least square estimator for the data points restricted to $I$. When $I = [\tilde{n}]$, we simplify \textproc{leastSquares}($\boldsymbol{X}, \boldsymbol{Y}, I$)  to \textproc{leastSquares}($\boldsymbol{X}, \boldsymbol{Y}$).  }
\end{definition}

Fig.~\ref{fig:\modelName} illustrates the structure of~\modelName. The depths of its leaf nodes may be different, \emph{i.e.},~\modelName~is an unbalanced tree.
Instead of having key-pointer pairs, a node in~\modelName~contains a model for indexing purpose.
Specifically, a node $\mathsf{N}$ keeps two numbers $\mathsf{N}.\mathtt{lb}$ and $\mathsf{N}.\mathtt{ub}$ such that $[\mathsf{N}.\mathtt{lb}, \mathsf{N}.\mathtt{ub})$ forms $\mathsf{N}$'s \textbf{range}, \emph{i.e.}, the key sequence covered by $\mathsf{N}$. 
A node $\mathsf{N}$, be internal or not, also stores a linear regression model $\mathsf{N}.\mathcal{LR}$ parameterized by its intercept $a$ and slope $b$, \emph{i.e.}, $\mathsf{N}.\mathcal{LR}(x) = a + bx$. Such models serve different purposes in internal and leaf nodes.

\noindent\underline{\textbf{Internal Nodes.}}~\modelName's internal nodes are represented as red dotted boxes in the bottom-middle part of Fig.~\ref{fig:\modelName}.
An internal node $\mathsf{N_T}$ stores a linear regression model $\mathsf{N_T}.\mathcal{LR}$ and an array $\mathsf{N_T}.\boldsymbol{C}$ of pointers to $\mathsf{N_T}$'s child nodes. 
Note that $\mathsf{N_T}.\boldsymbol{C}$'s each element is a simple pointer, without any keys.
In Fig.~\ref{fig:\modelName}, an internal node's child nodes are represented as small \textbf{equal sized} rectangles. This is because $\mathsf{N_T}$'s child nodes equally divide $\mathsf{N_T}$'s range.
Unlike B+Tree, we impose no constraints on $\mathsf{N_T}$'s fanout, \emph{i.e.}, the length of $\mathsf{N_T}.\boldsymbol{C}$. 
Also, an internal node $\mathsf{N_T}$ in~\modelName~does not need to store an additional ordered set of elements to describe the children's ranges, because they are clearly described by the linear regression model.
Given a key $x$, we can easily know which child node covers $x$ with a few simple calculations.
When a search goes downward in the tree, $\mathsf{N_T}$ uses $\mathsf{N_T}.\mathcal{LR}$ to `compute' the location (in $\mathsf{N_T}.\boldsymbol{C}$) of the pointer to the next child node to visit.
Let $\mathsf{N_T}.\mathtt{fo}$ denote the fanout of $\mathsf{N_T}$. The intercept $a$ and the slope $b$ of $\mathsf{N_T}.\mathcal{LR}$ are calculated as follows:
\begin{flalign}\label{Eq:internal_lr}
	b &= \mathsf{N_T}.\mathtt{fo} / (\mathsf{N_T}.\mathtt{ub} - \mathsf{N_T}.\mathtt{lb}), a = -b \times \mathsf{N_T}.\mathtt{lb}
\end{flalign}

\noindent Accordingly, $\mathsf{N_T}$'s $i$th child node's range is $[\mathsf{N_T}.\mathtt{lb} + \frac{i}{b}, \mathsf{N_T}.\mathtt{lb} + \frac{i+1}{b}\big)$, \emph{i.e.}, $[\mathsf{N_T}.\mathcal{LR}^{-1}(i), \mathsf{N_T}.\mathcal{LR}^{-1}(i+1) \big)$. All $\mathsf{N_T}$'s child nodes' ranges are of equal length.
For example, in Fig.~\ref{fig:\modelName}, the internal node $\mathsf{N_T}$ has four children and its range is $[80, 160)$.  $\mathsf{N_T}$'s second child node $\mathsf{N_D^1}$ is assigned a range of $[\mathsf{N_T}.\mathtt{lb} + \frac{1}{\mathsf{N_T}.b}, \mathsf{N_T}.\mathtt{lb} + \frac{2}{\mathsf{N_T}.b}\big) = [100, 120)$.

\noindent \underline{\textbf{Leaf Nodes.}} The leaf nodes in~\modelName~are represented by circles in Fig.~\ref{fig:\modelName}, where the ellipse in the bottom-middle part gives the details. 
A leaf node $\mathsf{N_D}$ stores an \textbf{entry} array $\boldsymbol{V}$ and a linear regression model $\mathcal{LR}$. Each entry is a pair, a pointer to another leaf node or a NULL flag indicating the corresponding slot of $\boldsymbol{V}$ is empty. 
$\mathsf{N_D}.\boldsymbol{V}$ may cover leaf node pointers or NULL flags due to the \textbf{\emph{local optimization}} strategy applied to the leaf nodes, which will be introduced later in this section and Section~\ref{sec:optimization}.
The learned model $\mathcal{LR}$ maps a key to a position in $\boldsymbol{V}$.
Unlike those models in internal nodes, $\mathcal{LR}$ is the solver to the mean squared error minimization problem, whose input is the keys of $\boldsymbol{P}_{L}$, \emph{i.e.}, the pairs covered by $\mathsf{N_D}$'s range, and ground truth is the corresponding indices in $\boldsymbol{P}_{L}$. Specifically, $\mathsf{N_D}.\mathcal{LR} \triangleq \textproc{leastSqaures}(\textproc{keys}(\boldsymbol{P}_{L}), \tilde{[|\boldsymbol{P}_{L}|]})$. 
At present, we simply assume $\mathsf{N_D}.\boldsymbol{V} = \boldsymbol{P}_{L}$. In other words, $\boldsymbol{V}$ tightly stores the key-pointer pairs only.
Note that each internal or leaf node only needs to store two parameters $a$ and $b$ for its linear model.

\noindent \underline{\textbf{Search without Optimization.} }
To search for a key $x$,  we first find the leaf node whose range covers $x$, using the function \textproc{locateLeafNode} (lines~5--8 in Algorithm~\ref{alg:search}).
This function starts at~\modelName's root, iteratively uses the linear regression model in the current internal node to `compute' a location in the node's pointer array $\boldsymbol{C}$. It follows the pointers to child nodes iteratively until reaching a leaf node $\mathsf{N_D}$.
As the internal nodes' linear regression models have perfect accuracy, \emph{i.e.}, they always choose the child nodes covering $x$. Thus, no local search is needed inside an internal node.

\begin{smaller}
	\begin{algorithm}[htb]
		\caption{\textproc{Search}($\mathsf{Root}, x$)}
		\label{alg:search}
		\begin{algorithmic}[1]
			\smaller
			
			\State {$\mathsf{N_D} \gets$ \textproc{locateLeafNode}($\mathsf{Root}, x$) }
			\State{$pos' \gets \lfloor \mathsf{N_D}.\mathcal{LR}(x) \rfloor$}
			\State {$p \gets \mathsf{N_D}.\boldsymbol{V}[pos]$, $pos \gets$ \textproc{exponentialSearch}($\mathsf{N_D}.\boldsymbol{V}, x, pos'$)}
			
			\State{\textbf{return} ($p.key = x$ ? $p.ptr$ :  NULL)}
			
			\Function{locateLeafNode}{$\mathsf{Root}, x$}
			\While {$\mathsf{N}$ points to an internal node}
			\State{$pos \gets \lfloor \mathsf{N}.\mathcal{LR}(x) \rfloor$;~ $\mathsf{N} \gets \mathsf{N}.\boldsymbol{C}[pos]$}
			\EndWhile
			\State {\textbf{return} $\mathsf{N}$}
			\EndFunction
		\end{algorithmic}
	\end{algorithm}
\end{smaller}

After finding the leaf node $\mathsf{N_D}$, we search the pair array $\mathsf{N_D}.\boldsymbol{V}$ for the pair whose key is $x$ (lines~2--4).
Suppose the pair $p \in \mathsf{N_D}.\boldsymbol{V}$ is the \emph{least upper bound}\footnote{In a pair array $\boldsymbol{P}$, a key $x$'s \textbf{least upper bound (LUB)} is a pair $p \in \boldsymbol{P}$ satisfying two conditions: 1) $p.key \geq x$ and 2) If $\exists p' \in \boldsymbol{P}$ \emph{s.t.} $p'.key \geq x$, then $p'.key \geq p.key$.} of $x$ in $\mathsf{N_D}.\boldsymbol{V}$. We use the model $\mathsf{N_D}.\mathcal{LR}$ to estimate $p$'s position $pos'$ in $\mathsf{N_D}.\boldsymbol{V}$ (line~2).
From the position $pos'$, an exponential search (line~3) is performed to find the actual position of $p$.
At the returned position $pos$ is the pair $p$ (line~3). If $p.key$ is not $x$, we return NULL as no data record contains the key $x$. Otherwise, we return $p.ptr$ that points to $x$'s data (line~5).

\noindent \underline{\textbf{Construction.}}~\modelName~is built by considering the keys' distribution to reduce the expected lookup time of queries.
We propose a novel bulk loading algorithm to build~\modelName~in Section~\ref{sec:bulk_load}. The algorithm `learns' a good node layout for~\modelName~from a given pair set $\boldsymbol{P}$.

\noindent \underline{\textbf{Local Optimization strategy.}} In the experiments, we find that the `last-mile search' in the leaf nodes (line~3 in Algorithm~\ref{alg:search}) is usually a bottleneck of the entire query. While our bulk loading algorithm make keys covered by a leaf node $\mathsf{N_D}$ almost linearly distributed, $\mathsf{N_D}.\mathcal{LR}$ cannot guarantee perfect accuracy. To address this issue, we put a local optimization on each leaf node after its range and linear regression model is determined. Inspired by the novel idea of LISA~\cite{DBLP:conf/sigmod/Li0ZY020} that \textbf{uses ML models to directly determine keys' storage positions instead of approximating them}, the local optimization makes~\modelName~avoid local search by forcibly placing pairs at the returned position by the linear regression model. 
If $pos = \mathsf{N_D}.\mathcal{LR}(p.key)$, the pair $p$ will be put at $ \mathsf{N_D}.\boldsymbol{V}[pos]$. If the predictions of multiple pairs by $\mathsf{N_D}.\mathcal{LR}$ are the same, \emph{i.e.}, they \emph{conflict}, a new leaf node will be created to deal with them. In this case, the original leaf node itself will have its own child node.

It is noteworthy that Algorithm~\ref{alg:search} is only used in~\modelName's bulk loading stage. In practice, a search algorithm with the local optimization will be adopted. The details of the local optimization as well as the optimized search algorithm are to be introduced in Section~\ref{sec:optimization}.

\noindent \underline{\textbf{Updates.}} ~\modelName~supports data updates. Our insertion algorithm will create new leaf nodes to cover conflicting pairs if insertions incur conflicts. Meanwhile, pairs covered by a leaf node will be redistributed when too many node creations degrades the search performance.
The details are to be given in Section~\ref{sec:data_update}.

\noindent \underline{\textbf{Discussion.}} 
As described above and illustrated in Fig.~\ref{fig:\modelName},~\modelName's internal and leaf nodes feature different structures. Their local arrays keep different types of elements, due to their different roles in search.
Internal nodes' role in the search process is to efficiently locate the leaf node covering the search key. To fulfill this, an internal node $\mathsf{N_T}$'s children nodes are assigned equal-size ranges through its model $\mathsf{N_T}$.$\mathcal{LR}$. Thus, $\mathsf{N_T}.\boldsymbol{V}$ arranges $\mathsf{N_T}$'s child nodes tightly.
In contrast, the search process needs to find the pair from a leaf node $\mathsf{N_D}$'s entry array. However, $\mathsf{N_D}.\mathcal{LR}$ may predict for multiple keys the same position in $\mathsf{N_D}.\boldsymbol{V}$. To process the conflicts, the local optimization is adopted to create new leaf nodes to store the conflicting keys. 
Since multiple keys may conflict at the same predicted position and at least one slot is preserved for each pair, it is possible that some slots in $\mathsf{N_D}.\boldsymbol{V}$ have no contents. Thus, unlike $\mathsf{N_T}.\boldsymbol{C}$, $\mathsf{N_D}.\boldsymbol{V}$ accommodates different kinds of elements.


\section{Search Cost Analysis}
\label{sec:cost_analysis}

This section is a preparation for the bulk loading algorithm in Section~\ref{sec:bulk_load}. We make a detailed cache-aware cost analysis of Algorithm~\ref{alg:search}. 
Note that no local optimization is assumed at present.
In other words, given any leaf node $\mathsf{N_D}$, $\mathsf{N_D}.\boldsymbol{V}$ contains no leaf node pointer and $\mathsf{N_D}.\mathcal{LR}$ does not guarantee perfect prediction accuracy.

\noindent \textbf{Cost Analysis.} Algorithm~\ref{alg:search} consists of two steps: 1) finding the leaf node covering the search key and 2) local search inside the leaf node. 
Given a pair $p$ with key $x$, suppose $\mathsf{N_D}$ (with depth $D$) is the leaf node covering $x$. Let $\mathsf{N_D}.\mathcal{LR}$'s prediction error for $x$ be $\epsilon_{x} = |\mathsf{N_D}.\mathcal{LR}(x) - pos|$ where $pos$ is $p$'s position in $\mathsf{N_D}.\boldsymbol{V}$. The estimated search cost of key $x$ is denoted by $\boldsymbol{\mathrm{T_{s}}}(x)$ as follows.
\begin{flalign} \label{Eq:TS}
	\boldsymbol{\mathrm{T_{s}}}(x) &\approx \big((D - 1)\times \boldsymbol{\mathrm{T_{is}}}(x) \big) + \boldsymbol{\mathrm{T_{ds}}}(\mathsf{N_D}, x) \\ \nonumber
	\boldsymbol{\mathrm{T_{is}}}(x) &= (\theta_{\mathsf{N}} + \eta + \theta_{\mathsf{C}}) ,
	\boldsymbol{\mathrm{T_{ds}}}(\mathsf{N_D}, x) = \theta_{\mathsf{N}} + \eta + \boldsymbol{\mathrm{t_{E}}}(\mathsf{N_D}, x)
\end{flalign}

\noindent where $\boldsymbol{\mathrm{T_{is}}}(x)$ and $\boldsymbol{\mathrm{T_{ds}}}(\mathsf{N_D}, x)$ denote the time spent in an internal node and the leaf node $\mathsf{N_D}$ covering $x$ respectively;
$\theta_{\mathsf{N}}$ and $\eta$ are the estimated time of executing a linear function (lines 2 and 8 in Algorithm~\ref{alg:search}) and loading a~\modelName's node from the main memory respectively;
$\theta_{\mathsf{C}}$ is the estimated time of accessing the address of an internal node $\mathsf{N}$'s child node.
In particular, after calculating $pos = \lfloor \mathsf{N}.\mathcal{LR}(x) \rfloor$ (line 2 in Algorithm~\ref{alg:search}), we need to get the $pos$-th element from $\mathsf{N}.\boldsymbol{C}$, and $\theta_{\mathsf{C}}$ is the time of getting the corresponding pointer.
Usually, both $\theta_{\mathsf{N}}$ and $\theta_{\mathsf{C}}$ equal the time of loading a cache line sized block from the main memory to the cache.

An exponential search needs about $2\log_{2} \epsilon_{x} $ iterations.
Each iteration consists of the calculation of the middle position, an operation of pair addressing and a comparison of two keys. Thus, the estimated time of the local search in $\mathsf{N_D}$ is $\boldsymbol{\mathrm{t_{E}}}(\mathsf{N_D}, x) = 2 \log_{2} \epsilon_{x} \times (\mu_{\mathsf{E}} + \theta_{\mathsf{E}})$, where $\theta_{\mathsf{E}}$ and $\mu_{\mathsf{E}}$ are the average time of accessing a pair and executing the other operations in one iteration, respectively.
An exponential search fetches pairs mostly stored separately.

\noindent \textbf{Discussion.}
In practice, pair access is much slower than other operations. Due to limited cache size, a new node or pair often triggers a cache miss, which entails addressing in the heap. Addressing takes two steps: finding the leaf node and local search inside the node. However, less local search time in a leaf node often means the node stores fewer pairs, which tends to increase the number of leaf nodes. All this means more and deeper leaf nodes, which in turn incurs more time cost for finding the correct leaf node.
To strike a trade-off between~\modelName's leaf node depth and number of leaf nodes, we proceed to design a bulk loading method to construct~\modelName~by taking into account the time cost of both steps together.


\section{Construction of~\modelName}
\label{sec:bulk_load}

For a pair array $\boldsymbol{P}$ sorted on all keys, we want to build~\modelName~with a good node layout having fast lookups for arbitrary search keys.

\subsection{Motivation and Overall Idea of BU-Tree}
\label{subsec:overall_idea}

As described in Section~\ref{sec:structure},~\modelName's linear regression model in an internal node $\mathsf{N_T}$ has perfect accuracy because its children equally divide its range by design. This design gives rise to a unique critical problem in constructing~\modelName: deciding the suitable fanout for $\mathsf{N_T}$.

One idea is to follow the top-down construction of ALEX~\cite{DBLP:conf/sigmod/DingMYWDLZCGKLK20}, using a power of 2 for a internal node's fanout.
If the whole key range is $[0, 1)$, the length of a leaf node's range must be $\frac{1}{2^{k}}$ for some integer $k$.
For a complex key distribution (\emph{e.g.}, a long-tail type), $k$ must be large in order to ensure high accuracy of the linear model in the leaf node. This tends to result in many leaf nodes and thus a high tree, making it slow to find the leaf node for a given key.
	
Another idea comes from the bottom-up bulk loading of B+Tree.
First, we partition all pairs in $\boldsymbol{P}$ into pieces and store each piece in a leaf node $\mathsf{N_D}$.
In each $\mathsf{N_D}$, we build a linear model to map $\mathsf{N_D}$'s keys to their positions in $\mathsf{N_D}$'s piece.
With an appropriate algorithm, we can ensure a relatively low total loss of all linear models in all leaf nodes.
Then, we partition the boundaries of the leaf nodes using the same algorithm.
We create internal nodes at height level 1 to save the boundaries, use the boundaries as the separation values to group the leaf nodes, and make each group of leaf nodes the children of their corresponding parent node at level 1.
Likewise, we create internal nodes at height level $i$ based on those at level $i - 1$, and repeat the process until we reach a level with only one node. This approach reduces the local search time in the leaf nodes but it does not guarantee that child nodes equally divide their parent node's range, and thus the parent node's linear regression model fail to give perfect prediction accuracy. Rather, from the predicted position, extra operations must be performed to find the child node covering a given key, making the overall lookup time longer.

\begin{figure}[htb]
	\centering
	\includegraphics[width=0.47\textwidth, trim={5mm 5mm 5mm 5mm},clip]{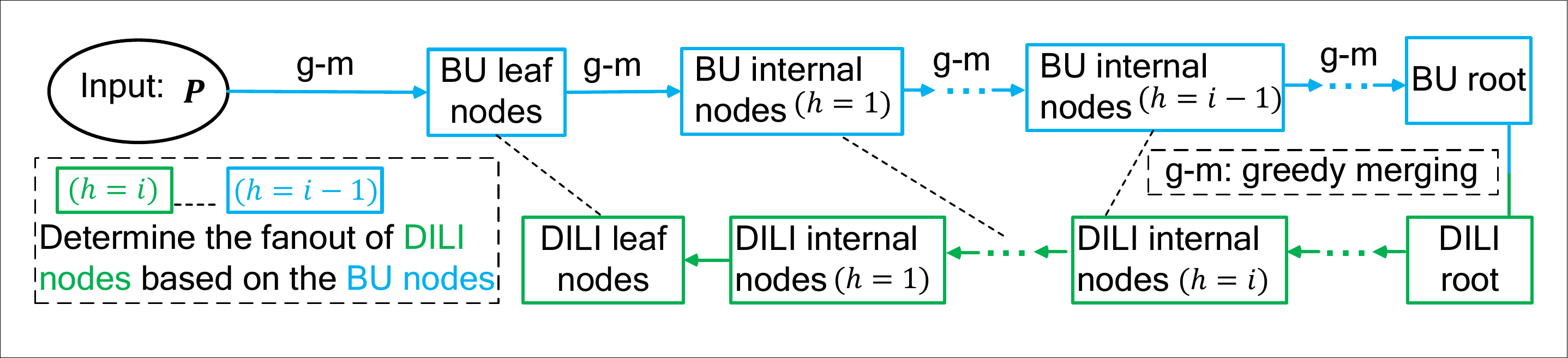}
	\caption{Framework of the bulk loading algorithm}
	\label{fig:DILI_construct}
\end{figure}

To build~\modelName~that incurs low overall lookup time for arbitrary keys, we combine both ideas in a two-phase bulk loading algorithm. 
First, we create a bottom-up tree (BU-Tree), starting from leaf nodes and growing the tree upwards.
Second, we reuse the BU-Tree's node layout to build~\modelName, and improve the latter's internal linear  models such that they also obtain perfect prediction accuracy.
The built~\modelName~is able to find the leaf node covering a search key with only a few calculations of linear functions. Also, the local search time cost in the leaf nodes is small as~\modelName~has a leaf node layout similar to the BU-Tree.
Thus, ~\modelName~is built in a novel paradigm---we first design a mirror model that finds leaf nodes efficiently, and then create a similar model that further optimizes local search in a leaf node.
Fig.~\ref{fig:DILI_construct} illustrates the procedure of our bulk loading algorithm.

We use \emph{BU internal node} or \emph{BU leaf node} to refer to an internal or leaf node in the BU-tree.
A BU internal node $\mathsf{N_T}$ is structurally the same as that in~\modelName, except that $\mathsf{N_T}$ stores an additional array $\boldsymbol{B}$ to record the ranges of all its child nodes.
Specifically, $\mathsf{N_T}.\boldsymbol{C}[i-1].\mathtt{ub} = \mathsf{N_T}.\boldsymbol{C}[i].\mathtt{lb} = \mathsf{N_T}.\boldsymbol{B}[i]$. Note that the child nodes may not equally divide $\mathsf{N_T}$'s range.
All BU leaf nodes are at the same height level and they are reused as the basis of the leaf nodes in~\modelName.

Key search in a BU-Tree is different from~\modelName.
Finding the child node covering key $x$ in a BU internal node $\mathsf{N_T}$ involves two steps. The first step computes $j = \mathsf{N_T}.\mathcal{LR}(x)$ and the second step searches $\mathsf{N_T}.\boldsymbol{B}$ from position $j$ to find the index $i$ such that $\mathsf{N_T}.\boldsymbol{B}[i] \leq x < \mathsf{N_T}.\boldsymbol{B}[i+1]$. As a result, $\mathsf{N_T}.\boldsymbol{\mathsf{C}}[i]$ points to the correct child node.

\subsection{Building BU-Tree}
\label{subsec:build_butree}

For a given $\boldsymbol{P}$, BU-Tree is built by Algorithm~\ref{alg:build_bottom_up}.
After initialization (line~1), it calls the function \textproc{greedyMerging} (Algorithm~\ref{alg:greedy_merge} to be detailed in Section~\ref{sssec:node_layout}) to generate all leaf nodes (line~2). Subsequently, Algorithm~\ref{alg:build_bottom_up} creates all BU internal nodes (lines~3--11) in a bottom-up way, until an appropriate root node is found (lines~7--9).
At each height $h$, we independently decide if the nodes at the current height should be the children of an immediate root node or not.
For both cases (lines~5~and~6), we calculate the average \emph{estimated accumulated search cost} (to be detailed in Section~\ref{sssec:node_layout}). It is an estimate of the lookup time of the corresponding~\modelName~from its root node to the node at height $h$ that covers the search key.
If having an immediate root node implies a smaller cost, we create a root node and set its child nodes to be $\boldsymbol{\mathsf{N}}^{h}$, the BU nodes at height $h$ (lines~7--9). Otherwise, the BU-Tree grows to height $h + 1$ (line~10).

\begin{algorithm}[htb]
	\smaller
	\caption{\textproc{BuildBUTree}($\boldsymbol{P}$)}
	\label{alg:build_bottom_up}
	\begin{algorithmic}[1]
		\smaller
		\State {$N \gets |\boldsymbol{P}|, \boldsymbol{X} \gets [x_{0},\cdots,x_{N-1}]$ where $x_{i} = \boldsymbol{P}[i].key$}
		\Statex \LeftComment{0} {Generate BU leaf nodes}
		\State {$n_{0}, \boldsymbol{X}_{0}, \boldsymbol{\mathsf{N}}^{0},
			\varepsilon^{1} \gets$ \textproc{greedyMerging}(NULL, $\boldsymbol{X}$)}

		\State {$h \gets 0$}
		\While {$n_{h} > 1$}
		
		\State {$\mathsf{N}^{r}, \varepsilon^{0} \gets $ \textproc{generateRoot}($\boldsymbol{\mathsf{N}}^{h}, \boldsymbol{X}_{h},  \boldsymbol{X}$) }

		\State {$n_{h+1}, \boldsymbol{X}_{h+1}, \boldsymbol{\mathsf{N}}^{h+1}, \varepsilon^{1} \gets $ \textproc{greedyMerging}($\boldsymbol{\mathsf{N}}^{h}, \boldsymbol{X}_{h}$) }
		\If {$\varepsilon^{0} < \varepsilon^{1}$} \Comment{Growing~\modelName~will result in larger cost}
		\State {Set $\mathsf{N}^{r}$ to be the root node of the BU-Tree}
		\State {\textbf{break}}
		\EndIf
		\State{\textbf{else} $h \gets h + 1$}
		\EndWhile
		\State {\textbf{return} the root node}
		
		\Function{generateRoot}{$\boldsymbol{\mathsf{N}}^{h-1}, \boldsymbol{X}_{h-1}, \boldsymbol{X} $}
		
		\State {$\boldsymbol{Y}_{h-1} \gets I,  I \gets \tilde{[n_{h-1} ]}, n_{h-1} \gets |\boldsymbol{X}_{h-1} |$}
		\State {$\mathcal{F} \gets $  \textproc{leastSquares}($\boldsymbol{X}_{h-1}, \boldsymbol{Y}_{h-1}, I$)}
		\State {$\mathsf{R} \gets $ an empty BU internal node}
		\State {$\mathsf{R}.\mathcal{LR} \gets \mathcal{F}$, $\mathsf{R}.\mathtt{fo} \gets n_{h-1}, $ $\mathsf{R}.\boldsymbol{C} \gets \boldsymbol{\mathsf{N}}^{h-1}$}
		\State {$\varepsilon \gets \frac{1}{N} \sum_{i=0}^{N-1} \boldsymbol{\mathrm{T_{ns}^{B}}} (\mathsf{N}, x_{i})$} \Comment{Calculate the search cost}
		\State {\textbf{return} $\mathsf{R}$, $\varepsilon$}
		\EndFunction
	\end{algorithmic}
\end{algorithm}

\subsubsection{Bottom-up Node and Model Creation}
\label{sssec:nodes_creation}
Given the pair set $\boldsymbol{P}$, we have $\boldsymbol{X} = \textproc{keys}(\boldsymbol{P}) = [x_{0},\cdots,x_{N-1}]$ and $\boldsymbol{Y} = \tilde{[N]} = [0,\cdots,N-1]$ where $x_{i} = \boldsymbol{P}[i].key$.
We first find a suitable integer $n_{0}$ and $n_{0} - 1$ break points $[\beta_{1}^{0},\cdots,\beta_{n_{0}-1}^{0}]$ to partition the key space $\textproc{keys}(\boldsymbol{P})$ into $n_{0}$ pieces. The $i$th piece's range is equal to $[\beta_{i}^{0}, \beta_{i+1}^{0})$ where $\beta_{0}^{0} = \inf \textproc{keys}(\boldsymbol{P})$ and $\beta_{n_{0}}^{0} = \sup \textproc{keys}(\boldsymbol{P})$.
For the $i$th piece, supposing $\beta_{i}^{0} \leq x_{l} < \cdots < x_{r} < \beta_{i+1}^{0}$, we train a linear regression  model $\mathcal{F}_{i}^{0}$ with input $[x_{l},\cdots,x_{r}]$ and $[l,\cdots,r]$.
Then, $n_{0}$ BU leaf nodes are created. The $i$th node $\mathsf{N}_{i}^{0}$ is described as follows.

\begin{small}
	\begin{flalign}\label{Eq:bottom_up_data}
		\mathsf{N}_{i}^{0}.\mathtt{lb} &= \beta_{i}^{0}, \mathsf{N}_{i}^{0}.\mathtt{ub} = \beta_{i+1}^{0}, \mathsf{N}_{i}^{0}.\mathcal{LR}(x) = \mathcal{F}_{i}^{0}(x) - l, \mathsf{N}_{i}^{0}.\boldsymbol{V} = \boldsymbol{P}[l:r]
	\end{flalign}
\end{small}

Suppose that the BU nodes at height $h-1$ have been created. We define two lists $\boldsymbol{X}_{h-1} = [\mathsf{N}_{0}^{h-1}.\mathtt{lb}\cdots,\mathsf{N}_{n_{h-1}-1}^{h-1}.\mathtt{lb}]$ and $\boldsymbol{Y}_{h-1} = [0,\cdots,n_{h-1}-1]$, where $\mathsf{N}_{i}^{h-1}$ is the $i$th node at height $h-1$ and $n_{h-1}$ is the number of the nodes at height $h-1$. Similarly, we generate $n_{h} - 1$ break points $[\beta_{1}^{h},\cdots,\beta_{n_{h}-1}^{h}]$, partition the space into $n_{h}$ pieces, and build $n_{h}$ linear regression models. Given a key $x$, suppose $\mathsf{N}_{l}^{h-1}.\mathtt{lb} \leq x < \mathsf{N}_{l}^{h-1}.\mathtt{ub}$. We define a function $\zeta^{h-1}(x) = l$. The $i$th node $\mathsf{N}_{i}^{h}$ at height $h$ is described as follows.

\begin{small}
	\begin{flalign}\label{Eq:bottom_up_internal}
		\mathsf{N}_{i}^{h}.\mathtt{lb} &= \beta_{i}^{h}, \mathsf{N}_{i}^{h}.\mathtt{ub} = \beta_{i+1}^{h}, \mathsf{N}_{i}^{h}.\mathtt{fo} = n_{h}, \\ \nonumber
		\mathsf{N}_{i}^{h}.\mathcal{LR}(x) &= \mathcal{F}_{i}^{h}(x) - \zeta^{h-1}(x), \mathsf{N}_{i}^{h}.\boldsymbol{\mathsf{C}}[j] = \mathsf{N}_{q}^{h-1}, \\ \nonumber
		\mathsf{N}_{i}^{h}.\boldsymbol{B}[j] &= \mathsf{N}_{q}^{h-1}.\mathtt{lb}, \text{where~} q = \zeta^{h-1}
		(\beta_{i}^{h}) + j
	\end{flalign}
\end{small}

A key challenge here is to decide $n_{h}$ (the number of nodes at height $h$) and $[\beta_{1}^{h},\cdots,\beta_{n_{h}-1}^{h}]$ (the break points for these $n_{h}$ nodes), as they determine the node layout at height $h$.
\begin{figure*}[htb]
	\centering
	\includegraphics[width=0.98\textwidth,trim={5mm 6mm 5mm 7.5mm},clip]{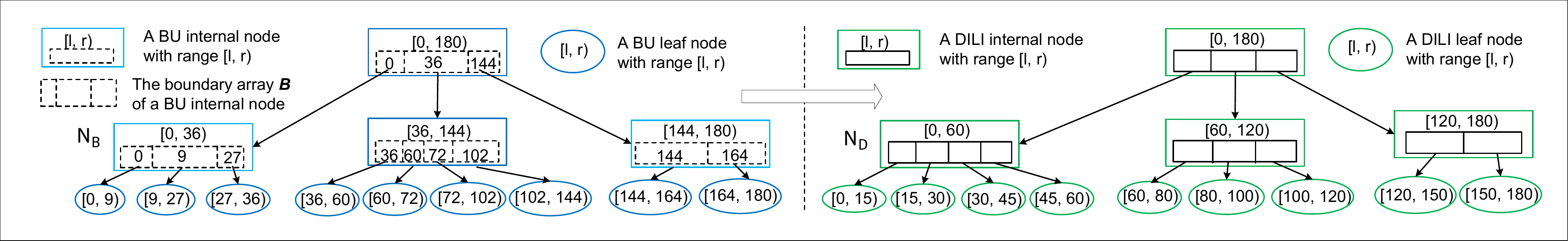}
	\caption{Building~\modelName~based on the BU-Tree}
	\label{fig:convert}
\end{figure*}

\subsubsection{Determining Node Layout at A Height}
\label{sssec:node_layout}

We want to have a suitable BU node layout such that the corresponding~\modelName~will have a good node layout to minimize the average search time. As the~\modelName~has a similar node layout with the BU-Tree, we simulate the~\modelName's querying process in the BU-Tree. To search for a key in the BU-Tree, we observe which nodes are accessed as well as the losses of their linear regression models. Based on those observations, we estimate the cost of searching for a key in the~\modelName.

Given a key $x$, we define the \emph{estimated search cost} $\boldsymbol{\mathrm{T_{ns}^{B}}}(\mathsf{N}, x, h)$ \emph{w.r.t.} a BU node $\mathsf{N}$ and a height $h$ as follows.
\begin{small}
\begin{flalign} \nonumber
&\boldsymbol{\mathrm{T_{ns}^{B}}}(\mathsf{N}, x, h) = \theta_{\mathsf{N}} + \eta + \rho^{h} \times \boldsymbol{\mathrm{t_{E}^{B}}}(\mathsf{N}, x), \text{where~} \rho \in (0, 1)\\ \label{Eq:bunode_lookup_cost}
&\boldsymbol{\mathrm{t_{E}^{B}}}(\mathsf{N}, x) = \log_{2} |\mathsf{N}.\mathcal{LR}(x) - i| \times (\mu_{\mathsf{E}} + \theta_{\mathsf{E}})\\ \nonumber
& i = \begin{cases}
	\mathsf{N}.\boldsymbol{B}[i] \leq x < \mathsf{N}.\boldsymbol{B}[i+1], &\mathsf{N} \text{~is a BU internal node}\\
	\mathsf{N}.\boldsymbol{V}[i].key \leq x < \mathsf{N}.\boldsymbol{V}[i+1].key, &\text{otherwise.}
\end{cases}
\end{flalign}
\end{small}
Here, $\theta_{\mathsf{N}}$,  $\eta$,  $\mu_{\mathsf{E}}$ and $\theta_{\mathsf{E}}$ carry the same meanings as those in Eq.~\ref{Eq:TS}; $\boldsymbol{\mathrm{t_{E}^{B}}}$ is a simple extension of $\boldsymbol{\mathrm{t_{E}}}$ on BU nodes.
If the node height is irrelevant, we simplify $\boldsymbol{\mathrm{T_{ns}^{B}}}(\mathsf{N}, x, h)$ to $\boldsymbol{\mathrm{T_{ns}^{B}}}(\mathsf{N}, x)$.

Given a key $x$, suppose the search in BU-Tree visits the complete node path $\mathsf{N}_{k}$, $\mathsf{N}_{k-1}$, $\cdots$, $\mathsf{N}_{0}$, where $\mathsf{N}_{k}$ is the root node and $\mathsf{N}_{0}$ is the leaf node covering $x$.
We define the \emph{accumulated search cost till height $h$} $\boldsymbol{\mathrm{T_{al}^{B}}}(h, x)$ and the \emph{complete search cost} $\boldsymbol{\mathrm{T_{s}^{B}}}(x)$ as follows.
\begin{small}
\begin{flalign} \nonumber
\boldsymbol{\mathrm{T_{al}^{B}}}(h, x) &= \sum_{j=h}^{k} \boldsymbol{\mathrm{T_{ns}^{B}}}(\mathsf{N}_{j}, x),
\boldsymbol{\mathrm{T_{s}^{B}}}(x) = \boldsymbol{\mathrm{T_{al}^{B}}}(0, x) = \sum_{i=0}^{k} \boldsymbol{\mathrm{T_{ns}^{B}}}(\mathsf{N}_{i}, x)
\end{flalign}
\end{small}

To minimize the average lookup time of the search key, we try to minimize $\frac{1}{N} \sum_{i=0}^{N-1} \boldsymbol{\mathrm{T_{s}^{B}}}(x_{i})$, \emph{i.e.}, the average complete search cost for all keys.
If the nodes under height $h$ have been created, minimizing $\frac{1}{N} \sum_{i=0}^{N-1} \boldsymbol{\mathrm{T_{s}^{B}}}(x_{i})$ is equivalent to minimizing $\frac{1}{N} \sum_{i=0}^{N-1} \boldsymbol{\mathrm{T_{al}^{B}}}(h, x_{i})$.
However, as a BU-Tree is grown upwards, we do not even know the height of the BU-Tree when creating nodes at height $h$, let alone estimating the search cost of a key in a node above height $h$.
To this end, we introduce the \emph{estimated accumulated search cost of the break points list $\boldsymbol{X}_{h}$} for the key set $\boldsymbol{X}$, termed as $\boldsymbol{\mathrm{T_{ea}^{B}}}(\boldsymbol{X}_{h}, \boldsymbol{X})$. It measures the quality of the node layout generated from the break points list $\boldsymbol{X}_{h}$. For simplicity, we assume that each BU internal node has the same number of child nodes, and define
$\boldsymbol{\mathrm{T_{ea}^{B}}}(\boldsymbol{X}_{h}, \boldsymbol{X})$ as follows.
\begin{small}
\begin{flalign} \nonumber
\boldsymbol{\mathrm{T_{ea}^{B}}}(\boldsymbol{X}_{h}, \boldsymbol{X}) &= \frac{1}{N} \sum_{i=0}^{N-1} \sum_{h'=h}^{\lceil \delta \rceil} \min(1, \delta + 1 - h') \times \boldsymbol{\mathrm{T_{ns}^{B}}} (\mathsf{N}_{t_{i}}^{h}, x_{i}, h') \\ \nonumber
\text{where~} t_{i} &= \zeta^{h}(x_{i}), \delta = \log_{\frac{n_{h-1}}{n_{h}}} n_{h-1},  n_{-1} = N
\end{flalign}
\end{small}

\noindent Above, $\mathsf{N}_{t_{i}}^{h}$ is the node at height $h$ whose range covers $x_{i}$. In other words, $\boldsymbol{X}_{h}[t_{i}] \leq x < \boldsymbol{X}_{h}[t_{i}+1]$. Moreover, $\delta$ is the estimated depth of the nodes at height $h$. We explain $\delta$ by an example. Suppose $n_{h-1} = 1000$ and $n_{h} = 100$, \emph{i.e.}, the number of nodes at height $h-1$ and $h$ are 1000 and 100, respectively. A node at height $h$ has $\frac{1000}{100}$ = 10 child nodes on average. The estimated $n_{h+1} = \frac{100}{10} = 10, n_{h+2} = 1$ and thus the root node's height is $h+2$. Because $\delta$ is the estimated depth of the nodes at height $h$, $\delta$ = $(h + 2) - h + 1$ = 3 = $\log_{\frac{1000}{100}} 1000$.
As $\delta$ may not exactly be an integer, we add a multiplication factor $\delta + 1 - h'$ $= \delta - \lfloor \delta \rfloor$ before $\boldsymbol{\mathrm{T_{ns}^{B}}} (\mathsf{N}_{t_{i}}^{h}, x_{i}, h')$ when $h' = \lceil \delta \rceil$.
Here, $\boldsymbol{\mathrm{T_{ns}^{B}}} (\mathsf{N}_{t_{i}}^{h}, x_{i}, h')$ is an estimate of $\boldsymbol{\mathrm{T_{ns}^{B}}} (\mathsf{N}_{t_{i}}^{h'}, x_{i})$.

Given $\boldsymbol{X}_{h-1}$ and $\boldsymbol{Y}_{h-1}$, to find the best $n_{h}$ and $\boldsymbol{X}_{h}$, a straightforward approach is to set $n_{h}$ to be different values. For each specific $n_{h}$, we solve a $n_{h}$-piecewise linear regression problem with input $\boldsymbol{X}_{h-1}$ and $\boldsymbol{Y}_{h-1}$, and compute the estimated accumulated search cost of this configuration.
Then, we choose the configuration with the smallest accumulated search cost. However, it is costly to directly solve a number of $k$-piecewise linear regression problems.

Instead, we adapt an efficient greedy merging algorithm~\cite{DBLP:conf/icml/AcharyaDLS16} to iteratively solve a series of $k$-piecewise linear regression problems with input $\boldsymbol{X}_{h-1}$ and $\boldsymbol{Y}_{h-1}$.
At each iteration, we generate $k-1$ break points and calculate the estimated accumulated search cost for them. Those break points induce the smallest cost form $\boldsymbol{X}_{h}$, the basis for creating the nodes at height $h$.
Algorithm~\ref{alg:greedy_merge} finds the suitable $n_{h}$ and $\boldsymbol{X}_{h}$ and generates the nodes at height $h$.
In lines~6--7, $\omega$ is a pre-defined average maximum fanout for~\modelName's nodes. In our implementation, we set $\omega$ to 4096 as a~\modelName~with good search performance cannot have too many nodes.

\begin{algorithm}[htb]
\smaller
\caption{\textproc{greedyMerging}($\boldsymbol{\mathsf{N}}^{h-1}, \boldsymbol{X}_{h-1}$) }
\label{alg:greedy_merge}
\begin{algorithmic}[1]
	\smaller
	\State {$k \gets \frac{n_{h-1}}{2}$, $\boldsymbol{Y}_{h-1} \gets \tilde{[n_{h-1}]}$, $n_{h-1} \gets |\boldsymbol{X}_{h-1} |$}
	\Statex \LeftComment{0} {The last set may contain 3 elements}
	\State {$\mathcal{I}^{k} \gets \{\{0, 1\}, \{2, 3\},\cdots,\{2k-2, (2k-1), n_{h-1} - 1\} \}$}
	\State {$\forall k \text{~and~} 0 \leq i < k$, let $I_{i}^{k}$ denote the $i$th element of $\mathcal{I}^{k}$} 
	\State {For any indices set $I$, let $\gamma(I) \gets \textproc{rmse}(\boldsymbol{X}_{h-1}, \boldsymbol{Y}_{h-1}, I)$}
	\State {$\forall k, i,$ let $s_{i}^{k} \gets \gamma(I_{i}^{k}), m_{i}^{k} \gets \gamma(I_{i}^{k} \bigcup I_{i+1}^{k})$}
	
	\State {Set $\omega$ to be a large number and $k_{min} \gets \frac{n_{h}}{\omega}$}  \Comment{In practice, we set $\omega = 2,048$.}

	\While {$k\geq k_{min}$} \Comment{Iterative greedy merging}
	\Statex \LeftComment{1}{$m_{i}^{k} = \gamma(I_{j}^{k} \bigcup I_{j+1}^{k}), s_{i}^{k} = \gamma(I_{j}^{k}), s_{i+1}^{k} = \gamma( I_{j+1}^{k})$}
	\State {$ u = \argmin_{i} m_{i}^{k} - s_{i}^{k} - s_{i+1}^{k}$}
	\State {$\mathcal{I}^{k-1} \gets \{I_{0}^{k}, I_{1}^{k},\cdots,I_{u}^{k}\bigcup I_{u+1}^{k},I_{u+2}^{k},\cdots,I_{k-1}^{k} \}$}
	\State {$s_{u}^{k-1} \gets m_{u}^{k}$ and calculate $m_{u - 1}^{k-1}$ and $m_{u}^{k-1}$}
	\State {$\forall u < i < k - 1, s_{i}^{k-1} \gets s_{i+1}^{k}, m_{i}^{k-1} \gets m_{i+1}^{k}$} 
	\State {$k \gets k - 1$}
	\Statex \LeftComment{1} {Generate new break points}
	\State {$\forall 0 \leq i < k, q_{i} \gets \inf I_{i}^{k}$}  
	\State {$\boldsymbol{B}_{k} = [\boldsymbol{X}_{h-1}[q_{0}], \boldsymbol{X}_{h-1}[q_{1}],\cdots,\boldsymbol{X}_{h-1}[q_{k-1}]]$}
	\State {$\varepsilon_{k} \gets \boldsymbol{\mathrm{T_{ea}^{B}}}(\boldsymbol{B}_{k}, \boldsymbol{X})$} \Comment{Get $\boldsymbol{B}_{k}$'s estimated accumulated search cost}
	\EndWhile
	\State {$n_{h} \gets \argmin_{k} \varepsilon_{k}$}
	\State {$\boldsymbol{X}_{h} \gets \boldsymbol{B}_{n_{h}}$}
	\State {$\forall 0 \leq i < n_{h}, \mathcal{F}_{i}^{h} \gets$ \textproc{leastSquares}($\boldsymbol{X}_{h-1}, \boldsymbol{Y}_{h-1}, I_{i}^{n_{h}}$)}
	\State {$\boldsymbol{\mathsf{N}}^{h} \gets$ the BU nodes at height $h$ described in Eq.~\ref{Eq:bottom_up_data} or Eq.~\ref{Eq:bottom_up_internal}}
	\State {\textbf{return} $n_{h}$, $\boldsymbol{X}_{h}$, $\boldsymbol{\mathsf{N}}^{h}$, $\varepsilon$}
\end{algorithmic}
\end{algorithm}

The value of $k$ is decided as follows. Initially, $k$ is set to be $\frac{n_{h-1}}{2}$ and the list $\boldsymbol{X}_{h-1}$ is partitioned into $k$ pieces  (lines~1-2).
At each iteration (lines~7--15), we merge two continuous pieces that result in the most linear loss increase and decrease the value of $k$ by 1 (lines~8-12).
We do not need to calculate the linear loss \emph{w.r.t.} every piece. Instead, we maintain two number $s_{i}^{k}$ and $m_{i}^{k}$ for the $i$th piece $I_{i}^{k}$. They equal the linear loss \emph{w.r.t.}  $I_{i}^{k}$ and $I_{i}^{k} \bigcup I_{i+1}^{k}$, respectively. Each iteration involves the change of one piece only. Thus, after deciding merging $I_{u}^{k}$ and $I_{u+1}^{k}$, we only need to update the values of $s_{u}^{k-1}$ and $m_{u}^{k-1}$ \emph{w.r.t.} $I_{u}^{k-1}$ (\emph{i.e.}, $I_{u}^{k} \bigcup I_{u+1}^{k}$) and $m_{u-1}^{k-1}$ \emph{w.r.t.} $I_{u-1}^{k-1}$. Clearly, $s_{u}^{k-1} = m_{u}^{k}$ and thus only two calculations of $m_{u-1}^{k-1}$ and $m_{u}^{k-1}$ are needed.
Meanwhile, the break points of the $k$-piecewise linear function $\boldsymbol{B}_{k}$ are generated (lines~13-14).
We compute $\boldsymbol{\mathrm{T_{ea}^{B}}}(\boldsymbol{B}_{k}, \boldsymbol{X})$ for each $k$ (line~13).
After the iterations, we set $\boldsymbol{X}_{h}$ to $\boldsymbol{B}_{n_{h}}$ and $n_{h}$ to the $k$ with the smallest $\boldsymbol{\mathrm{T_{ea}^{B}}}(\boldsymbol{B}_{k}, \boldsymbol{X})$ (lines~16-17).
The nodes at height $h$ are then created (lines~18--19) according to Eq.~\ref{Eq:bottom_up_data} and Eq.~\ref{Eq:bottom_up_internal}.

At each iteration, we calculate $m_{u-1}^{k-1}$ and $m_{u}^{k-1}$ to estimate the linear losses \emph{w.r.t.} two pieces. In implementation, we make the number of items in each piece smaller than a pre-defined threshold of $2\omega$. Thus, both calculations run in time $O(1)$.
We use a priority queue to store $d_{u}^{k} = m_{i}^{k} - s_{i}^{k} - s_{i+1}^{k}$ for all $i$. Thus, the time complexity of selecting $u$ (line~9) is $O(k) = O(n_{h})$. 
Besides, the calculation of the estimated accumulated search cost of $\boldsymbol{B}_{k}$ runs in time $O(1)$. In summary, the time complexity of Algorithm~\ref{alg:greedy_merge} is $O(n_{h}\log_{2} n_{h})$.

\subsection{BU-Tree based Bulk Loading for~\modelName}
\label{subsec:bulk_load}

\noindent Algorithm~\ref{alg:bulk_loading} formalizes bulk loading for~\modelName. In line~1, the BU-Tree is created by~\textproc{buildBUTree} (Algorithm~\ref{alg:build_bottom_up}).
Let $H$ be the BU-Tree's height. At any height $h \leq H$,~\modelName~and the BU-Tree have the same number of nodes, but the node layouts may be different. Based on the BU-Tree, we grow~\modelName~top down (lines~3--7). The range of~\modelName's root node $\mathsf{Root}$ is set to the counterpart in the BU-Tree. $\mathsf{Root}$ is created by the recursive function~\textproc{createInternal} (line~7).

\begin{algorithm}[htb]
\smaller
\caption{\textproc{BulkLoading}($\boldsymbol{P}$)}
\label{alg:bulk_loading}
\begin{algorithmic}[1]
	\smaller
	\State {$\mathsf{BURoot} \gets $ \textproc{buildBUTree}($\boldsymbol{P}$)}
	\State {$H \gets $ the height  of $\mathsf{BURoot}$}
	\State {Get $\boldsymbol{\mathsf{N}}^{0}, \boldsymbol{\mathsf{N}}^{1}, \cdots, \boldsymbol{\mathsf{N}}^{H-1}$ from $\mathsf{N}'$ }
	\For {$i \in \{0,1,\cdots,H-1\}$}
	\State {$\boldsymbol{\theta}^{i} = [\boldsymbol{\mathsf{N}}^{i}[0].\mathtt{lb}, \boldsymbol{\mathsf{N}}^{i}[1].\mathtt{lb}, \cdots, \boldsymbol{\mathsf{N}}^{i}[|\boldsymbol{\mathsf{N}}^{i}|-1].\mathtt{lb}] $}
	\EndFor
	\State {$\boldsymbol{\Theta} \gets [\boldsymbol{P}, \boldsymbol{\theta}^{0}, \boldsymbol{\theta}^{1}, \cdots, \boldsymbol{\theta}^{H-1}] $}
	\State {$\mathsf{Root} \gets $ \textproc{createInternal}($\mathsf{BURoot}.\mathtt{lb}, \mathsf{BURoot}.\mathtt{ub}, H, \boldsymbol{\Theta}$) }
	
	\State {\textbf{return} $\mathsf{Root}$}

	\Function {createInternal} {$\mathtt{lb}, \mathtt{ub}$, $h, \boldsymbol{\Theta}$}
	\State {$\mathsf{N_T} \gets $ an empty~\modelName~internal node}
	
	\State {$\mathsf{N_T}.\mathtt{lb} \gets \mathtt{lb}, \mathsf{N_T}.\mathtt{ub} \gets \mathtt{ub}$}
	\State {$\mathsf{N_T}.\mathtt{fo} \gets \#\{x| x \in \boldsymbol{\theta}, \mathtt{lb} \leq x < \mathtt{ub} \},  \boldsymbol{\theta} \gets \boldsymbol{\Theta}[h-1]$}
	\State {$\mathsf{N_T}.\mathcal{LR}(x) = a + bx$, $a \gets  -b \times \mathtt{lb}, b \gets \frac{\mathsf{N_T}.\mathtt{fo}}{\mathtt{ub}-\mathtt{lb}}$}
	\For {$i \in \{0,1,\cdots, \mathsf{N_T}.\mathtt{fo}-1\}$}
	\State {$l \gets \mathtt{lb} + \frac{i}{b}$, $u \gets \mathtt{lb} + \frac{i+1}{b}$} \Comment{The lower/upper bound}
	\If {$h$ is 1} \Comment {Child nodes are leaf nodes}
	\State {$\mathsf{N_T}.\boldsymbol{C}(i) = $ \textproc{createLeafNode}($l, u, \boldsymbol{\Theta}[0]$) } \Comment {$\boldsymbol{\Theta}[0]$ is $\boldsymbol{P}$}
	\EndIf
	\State {\textbf{else} $\mathsf{N_T}.\boldsymbol{C}(i) = $ \textproc{createInternal}($l, u, h - 1, \boldsymbol{\Theta}$) }
	\EndFor
	
	\State {\textbf{return} $\mathsf{N_T}$}
	\EndFunction
	
	\Function {createLeafNode} {$\mathtt{lb}, \mathtt{ub}$, $\boldsymbol{P}$}
	\State {$l \gets \argmin_{i} \boldsymbol{P}[i].key \geq \mathtt{lb}$, $u \gets \argmin_{i} \boldsymbol{P}[i].key \geq \mathtt{ub}$}
	\State {$M \gets |\boldsymbol{P}_{\mathsf{D}}|$,  $\boldsymbol{P}_{\mathsf{D}} \gets \boldsymbol{P}[l:u]$}
	\State {$\mathsf{N_D} \gets $ an empty~\modelName~leaf node, $\mathsf{N_D}.\Omega \gets M$}
	
	\State {$\mathsf{N_D}.\mathcal{LR} \gets $ \textproc{leastSquares}(\textproc{keys}($\boldsymbol{P}_{\mathsf{D}}$), $\tilde{[M]}$)}
	\State {\textproc{LocalOpt}($\mathsf{N_D}, \boldsymbol{P}_{\mathsf{D}}$)}
	
	\State {\textbf{return} $\mathsf{N_D}$}
	\EndFunction
\end{algorithmic}
\end{algorithm}

To create an internal node $\mathsf{N_T}$ (lines~9--19), we set its range according to the input bounds (line~11), its fanout to the number of BU nodes at height $h-1$ whose range is covered by $\mathsf{N_T}$'s range (line~12), and its linear regression model accordingly (line~13).
We recursively create $\mathsf{N_T}.\mathtt{fo}$ nodes and make them equally divide $\mathsf{N_T}$'s range (lines~14--18). These nodes compose $\mathsf{N_T}.\boldsymbol{C}$.

When creating a leaf node $\mathsf{N_D}$ (lines~20--26), we include in $\boldsymbol{P}_\mathsf{D}$ the pairs with keys from $\mathsf{N_D}$'s range (lines~21--22). The model $\mathsf{N_D}.\mathcal{LR}$ is trained with the input \textproc{keys}($\boldsymbol{P}_\mathsf{D}$) (lines~24). The function \textproc{LocalOpt} distributes the pairs to the entry array $\mathsf{N_D}.\boldsymbol{V}$ (line~25), performing a local optimization on $\mathsf{N_D}$. The details will be given in Section~\ref{sec:optimization}.

Fig.~\ref{fig:convert} exemplifies building~\modelName.
The $i$th internal nodes of BU-Tree and~\modelName~at height $h$ may have different fanouts.
For example, when $h = 1$, node $\mathsf{N_T^B}$ (in the BU-Tree) has 3 child nodes but the~\modelName~node $\mathsf{N_T^D}$'s fanout is 4, 
because $\mathsf{N_T^D}$'s range $[0, 60)$ covers the left boundaries of the first four BU leaf nodes' ranges.

\subsection{Remarks}
In Section~\ref{sssec:node_layout}, $\boldsymbol{\mathrm{T_{ns}^{B}}}(\mathsf{N_{i}}, x)$ estimates $\boldsymbol{\mathrm{T_{is}}}(x)$ or $\boldsymbol{\mathrm{T_{ds}}}(\mathsf{N_{i}}, x)$ in Eq.~\ref{Eq:TS}, depending on if $\mathsf{N_{i}}$ is internal or not.
If $\mathsf{N_{i}}$ is a BU leaf node (\emph{i.e.}, $h = 0$), $\boldsymbol{\mathrm{T_{ns}^{B}}}(\mathsf{N_{i}}, x)$ and $\boldsymbol{\mathrm{T_{ds}}}(\mathsf{N_{i}}, x)$ are the same as long as $\mathsf{N_{i}.ifL}$ is true. 
Otherwise, $\boldsymbol{\mathrm{T_{ns}^{B}}}(\mathsf{N_{i}}, x)$ is not the same as $\boldsymbol{\mathrm{T_{is}}}(x)$. Our~\modelName~bulk loading (Algorithm~\ref{alg:bulk_loading}) makes the leaf node layouts of the BU-Tree and~\modelName~as alike as possible. However, lower accuracy of BU internal nodes' linear regression models would cause the two layouts to be more different. To this end, we modify $\boldsymbol{\mathrm{T_{ns}^{B}}}(\mathsf{N_{i}}, x)$ to strike a trade-off between the BU-Tree's height and the accuracy of its models. In addition, internal nodes at a higher height tend to have less impact on the layout of~\modelName's leaf nodes. Thus, we multiply $\boldsymbol{\mathrm{t_{E}^{B}}}(\mathsf{N_{i}}, x)$ by a factor $\rho^{h}$ in Eq.~\ref{Eq:bunode_lookup_cost} to reflect this effect.

Our cost models of BU-Tree and~\modelName~consider the effect of the cache locality and attempt to strike a good balance between the node fanouts and the tree height. 
In contrast, the B+Tree restricts its node fanout to a pre-defined range $[\frac{m}{2}, m]$. Thus, to find the next child node to visit in an internal node's child array, the B+Tree binary search needs to access the array $\lceil log_{2} m\rceil$ times in the worst case.
As $m$ is relatively large, the local search inside a B+Tree node often triggers many cache misses.
In contrast, finding the required child node in~\modelName~triggers only one cache miss.
Also, compared to B+Tree,~\modelName's internal nodes have larger average fanouts. Thus, it usually has a wider and shallower structure such that a search needs to traverse fewer nodes to locate the relevant leaf nodes.

Note that ALEX partitions the key space in a relatively static way as its node fanout is always a power of 2. This causes its linear models to have relatively low accuracy.
To this end, ALEX adopts a gapped array to increase the model accuracy in its leaf nodes. 
Still, this cannot guarantee optimal model accuracy---ALEX needs more time and triggers more cache misses than~\modelName, as shown in Section~\ref{sec:experiments}.
Unlike ALEX, our local optimization (to be detailed in Section~\ref{sec:optimization}) makes the models in~\modelName's leaf nodes perfectly accurate and thus shortens~\modelName's search time. Nevertheless,~\modelName~still outperforms ALEX even without the local optimization.


\section{Local Optimization of~\modelName}
\label{sec:optimization}

A leaf node $\mathsf{N_D}$'s linear model $\mathsf{N_D}.\mathcal{LR}$ approximates the relation between keys and their positions in the array $\mathsf{N_D}.\boldsymbol{V}$.
However, no model can always make perfect predictions. Our experiments imply that the exponential search in the leaf nodes often forms a bottleneck. 
Also, the leaf node structure does not consider insertions. On average, half pairs covered by a leaf node needs to be shifted for a single insertion. To address these issues, we propose a local optimization for~\modelName's leaf nodes. It helps the linear models make 100\% accurate predictions and avoid element shifting in insertions.

\begin{algorithm}[htb]
	\smaller
	\caption{\textproc{LocalOpt}($\mathsf{N_D}, \boldsymbol{P}_{\mathsf{D}}$)}
	\label{alg:LocalOpt}
	\begin{algorithmic}[1]
		\smaller
		\State {$N \gets |\boldsymbol{P}_{\mathsf{D}}|, \boldsymbol{X} \gets [x_{0},\cdots,x_{N-1}]$ where $x_{i} = \boldsymbol{P}_{\mathsf{D}}[i].key$}
		\State {$\mathsf{N_D}.\Delta \gets 0$, $\mathsf{N_D}.\texttt{fo} \gets \eta \mathsf{N_D}.\Omega (\eta > 1)$}
		\State {$\mathtt{fo}\gets \mathsf{N_D}.\mathtt{fo}$ and allocate $\mathsf{N_D}.\boldsymbol{V}$ with $\mathtt{fo}$ NULLs} 
		\State{Define $f_{\mathsf{D}}(x) \triangleq \min(\mathtt{fo}-1, \max(0, \mathsf{N_D}.\mathcal{LR}(x))$}
		\State{$\forall$ $i = 0$ to $N-1$, set $\boldsymbol{P}_{\mathsf{D}}^{i}$ an empty list }
		\For {$i \in \{ 0,1,\cdots, N-1\}$}
		\State{$t \gets f_{\mathsf{D}}(x_{i}), \boldsymbol{P}_{\mathsf{D}}^{t}$.append($\boldsymbol{P}_{\mathsf{D}}[i]$)}
		\EndFor
		\For {$t \in \{0, 1, \cdots, \mathtt{fo}-1 \}$}
		\If {|$\boldsymbol{P}_{\mathsf{D}}^{t}| = 1$}
		\State {$\mathsf{N_D}.\boldsymbol{V}[t] \gets \boldsymbol{P}_{\mathsf{D}}^{t}[0]$, $\mathsf{N_D}.\Delta +$= 1}
		\ElsIf{|$\boldsymbol{P}_{\mathsf{D}}^{t}| > 1$}
		\State {Create a new leaf node $\mathsf{N}'$ and train $\mathsf{N}'.\mathcal{LR}$ with the input $\boldsymbol{P}_{\mathsf{D}}^{t}$}
		\State {\textproc{LocalOpt}($\boldsymbol{P}_{\mathsf{D}}^{t}, \mathsf{N}')$}
		\State {$\mathsf{N_D}.\boldsymbol{V}[t] \gets $ the pointer to $\mathsf{N}'$, $\mathsf{N_D}.\Delta$ += |$\boldsymbol{P}_{\mathsf{D}}^{t}$| + $\mathsf{N}'.\Delta$}
		\EndIf
		\State {\textbf{else} $\mathsf{N_D}.\boldsymbol{V}[t] \gets$ NULL}
		\EndFor
		\State {$\mathsf{N_D}.\kappa  = \frac{\mathsf{N_D}.\Delta}{\mathsf{N_D}.\Omega}$}
	\end{algorithmic}
\end{algorithm}

\begin{figure}[htb]
	\centering
	\includegraphics[width=0.45\textwidth,trim={5mm 6mm 5mm 6mm},clip]{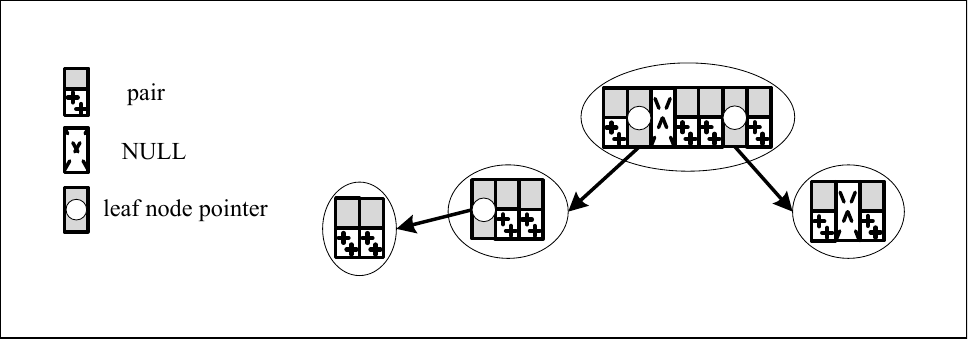}
	\caption{Leaf node structure}
	\label{fig:leaf_node_example}
\end{figure}

Our local optimization (Algorithm~\ref{alg:LocalOpt}) starts at the final step in the procedure of creating a leaf node $\mathsf{N_D}$ (line~25 in Algorithm~\ref{alg:bulk_loading}). Suppose there is one and only one pair whose predicted position by $\mathsf{N_D}.\mathcal{LR}$ is $t$, we set $\mathsf{N_D}.\boldsymbol{V}[t]$ to be this pair (lines 6-7, 9-10). If multiple pairs conflict at position $t$, we create a new leaf node $\mathsf{N}'$ to cover them (lines~11-12). The \textproc{LocalOpt} function is recursively called such that the entry array $\mathsf{N}'.\boldsymbol{V}$ is created to organize the conflicting pairs (line~13). After that, the pointer to $\mathsf{N}'$ is assigned to$\mathsf{N_D}.\boldsymbol{V}[t]$.
Those slots of $\mathsf{N_D}.\boldsymbol{V}$ without a pair are set to NULL (lines~15). 
In practice, we set $\mathsf{N_D}.\mathtt{fo}$ to  $\eta\cdot \mathsf{N_D}.\Omega$, where $\eta$ is an enlarging ratio, such that continuous keys are more likely assigned in different slots (line~2 ) and conflicts are reduced.
Fig.~\ref{fig:leaf_node_example} illustrates the structure of a typical leaf node.

The local optimization renders search via~\modelName~slightly different from Algorithm~\ref{alg:search}, as shown in Algorithm~\ref{alg:search_w_opt}. After finding the highest leaf node $\mathsf{N_D}$ covering $x$ (line~1), the loop starts from obtaining the value of the returned element $p$ in $\mathsf{N_D}.\boldsymbol{V}$ through $\mathsf{N_D}.\mathcal{LR}$.  If $p$ points to another leaf node $\mathsf{N}'$, the loop continues by setting $\mathsf{N_D}$ to $\mathsf{N}'$ (lines~5-6). Otherwise, the loop ends and returns the right result, depending on if $p$ is a pair with key equal to $x$ (lines~7-9).

\begin{algorithm}[htb]
	\smaller
	\caption{\textproc{SearchWOpt}($\mathsf{Root}, x$)}
	\label{alg:search_w_opt}
	\begin{algorithmic}[1]
		\smaller
		\State {$\mathsf{N_D} \gets$ \textproc{locateLeafNode}($\mathsf{Root}, x$) }
		\While{True}
		\State{$pos \gets \lfloor \mathsf{N_D}.\mathcal{LR}(x) \rfloor$}
		\State{$p \gets \mathsf{N_D}.\boldsymbol{V}[pos]$}
		\If{$p$ points to a leaf node $\mathsf{N}'$}
		\State{$\mathsf{N_D} \gets \mathsf{N}'$}
		\ElsIf{$p \neq$ NULL and  $p.key == x$}
		\State{\textbf{return} $p.key$}
		\EndIf
		\State{\textbf{else return} NULL} 
		\EndWhile
	\end{algorithmic}
\end{algorithm}

Fig.~\ref{fig:\modelName} gives an example of search via locally optimized~\modelName~for a key $x = 101$. The root node $\mathsf{R}$'s key range is $[0, 240)$. $\mathsf{R}$ has three child nodes. It is easy to derive that $\mathsf{R}.a = 0$, $\mathsf{R}.b = \frac{1}{80}$ such that $\mathsf{R}'s$ linear model equally divides its key range into three parts. By a simple calculation $\lfloor \mathsf{R}.a  + \mathsf{R}.b \times x \rfloor = 1$, we know $x$ is covered by $\mathsf{R}$'s second child node, namely $\mathsf{N_T}$ in Fig.~\ref{fig:\modelName} (Step-1).
$\mathsf{N_T}$ equally divides its range $[80, 160)$ to its four children with $\mathsf{N_T}.a = -4$ and $\mathsf{N_T}.b = 0.05$. The search goes into $\mathsf{N_T}$'s second child node $\mathsf{N_D^1}$ whose range is $[100, 120)$ (Step-2).
$\mathsf{N_D^1}$ is a leaf node whose linear model is trained with the keys covered by its range, in a different way from that of $\mathsf{R}$ and $\mathsf{N_T}$. At the predicted slot position $\lfloor \mathsf{N_D^1}.a  + \mathsf{N_D^1}.b \times x \rfloor = \lfloor -9  + 0.1 \times 101 \rfloor = 1$ is another leaf node $\mathsf{N_D^2}$ (Step-3). 
Note that two keys $101$ and $102$ conflict at the same slot in $\mathsf{N_D^1}.\boldsymbol{V}$. Thus, $\mathsf{N_D^2}$ is generated to store them as the local optimization. 
Finally, $\mathsf{N_D^2}.\mathcal{LR}$ predicts for $x$ a pair $p$ at position 0 in $\mathsf{N_D^2}.\boldsymbol{V}$. The output pointer $p_x = p.ptr$ points to the data record identified by $x$.


\section{Data Updates in~\modelName}
\label{sec:data_update}

\subsection{Insertions}
\label{subsec:insertion}

Insertions via~\modelName~are logically simple and efficient. Our insertion algorithm  avoids element shifting that happen to B+Tree and ALEX, and it redistributes pairs when insertions degrade the search performance. The details are shown in Algorithm~\ref{alg:insert}.

To insert a pair $p$ to~\modelName, the first step calls the function \textproc{locateLeafNode} (defined in Algorithm~\ref{alg:search}) to find the leaf node $\mathsf{N_D}$ that supposedly covers $p.key$ (line~1). 
Next, the algorithm inserts $p$ into $\mathsf{N_D}$ by calling the recursive function \textproc{insertToLeafNode} (line~2). We use the model $\mathsf{N_D}.\mathcal{LR}$ to calculate the position $pos$ in $\mathsf{N_D}.\boldsymbol{V}$ for $p$. Suppose that at position $pos$ is the element $p'$ (line~4).
If $p'$ is NULL, the $pos$-th slot of $\mathsf{N_D}.\mathcal{LR}$ is empty. We simply place $p$ at the slot. Then, the cost of searching for all pairs except for $p$ covered by $\mathsf{N_D}$ does not change. Searching for $p$ from $\mathsf{N_D}$ needs only one extra entry access, so we simply add one to $\mathsf{N_D}.\Delta$ (lines~6-7).
Here, $\mathsf{N_D}.\Delta$ denotes the total number of entries to be accessed to search for all keys covered by $\mathsf{N_D}$, starting from $\mathsf{N_D}$.
If $p'$ points to another leaf node $\mathsf{N'}$, we insert $p$ into $\mathsf{N}'$'s entry array (line~10). This time the change of $\mathsf{N_D}.\Delta$ is related to those pairs in $\mathsf{N}'$. Thus, we record $\mathsf{N}'.\Delta$ before the insertion (line~9). The increment of $\mathsf{N_D}.\Delta$ is the change of $\mathsf{N}'.\Delta$ plus 1 (line~11).
If $p$ exists (line~12), we do nothing (line~13).
If a conflict happens (line~14), we need to replace the pair $p'$ with a new leaf node covering $p$ and $p'$ at the $pos$-th position of $\mathsf{N_D}.\boldsymbol{V}$ (lines~15-17). In this case, $\mathsf{N_D}.\Delta$ is increased by three (line~18): one for searching for $p'$ and two for $p$.

\begin{algorithm}[htb]
	\smaller
	\caption{\textproc{Insert}($\mathsf{Root}, p$)}
	\label{alg:insert}
	\begin{algorithmic}[1]
		\smaller

		\State {$\mathsf{N_D} \gets$ \textproc{locateLeafNode}($\mathsf{Root}, p.key$) }
		\State {\textbf{return} \textproc{insertToLeafNode}($\mathsf{N_D}, p$)}
		
		\Function {insertToLeafNode} {$\mathsf{N_D}, p$}
		\State {$p' \gets \mathsf{N_D}.\boldsymbol{V}[pos]$, $pos \gets \mathsf{N_D}.\mathcal{LR} (p.key$)}
		\State {$\mathtt{notExist} \gets$ True}
		\If {$p' =$ NULL}
		\State {$\mathsf{N_D}.\boldsymbol{V}[pos] \gets p$, $\mathsf{N_D}.\Delta$ += 1 }  \Comment{insert $p$ to an empty slot}
		\ElsIf{$p'$ points to another leaf node $\mathsf{N}'$}
		\State {$\Delta' \gets \mathsf{N}'.\Delta$}
		\State {$\mathtt{notExist} \gets$ \textproc{insertToLeafNode}($\mathsf{N}', p$)}
		\State {$\mathsf{N_D}.\Delta$ += $1 + \mathsf{N'}.\Delta - \Delta'$}
		\ElsIf{$p'.key$ = $p.key$}
		\State {$\mathtt{notExist} \gets$ True}  \Comment{$p$ exists}
		\Else
		\State {create a new leaf node $\mathsf{N}'$ to cover $p$ and $p'$} 
		\State {$\mathsf{N}'.\Delta \gets 2$, $\mathsf{N}'.\Omega\gets 2$ and train $\mathsf{N}'.\mathcal{LR}$}
		\State {$\mathsf{N_D}.\boldsymbol{V}[pos] \gets$ the pointer to $\mathsf{N}'$}
		\State {$\mathsf{N_D}.\Delta$ += $1 + \mathsf{N'}.\Delta$}
		\EndIf
		\State {$\mathsf{N_D}.\Omega$ += ($\mathtt{notExist} =$ True ? 1 : 0)}
		\If {$\mathtt{notExist} =$ True and $\frac{\mathsf{N_D}.\Delta}{\mathsf{N_D}.\Omega} > \lambda \times \mathsf{N_D}.\kappa$}
		\State {collect all pairs covered by $\mathsf{N_D}$ and store them in list $\boldsymbol{P}_{\mathsf{D}}$}
		\State {$\mathsf{N_D}.\mathtt{fo} \gets \mathsf{N_D}.\Omega \times r$, $r \gets \varphi$($\mathsf{N_D}.\alpha$), $\mathsf{N_D}.\alpha$ += 1}
		\State {$\mathsf{N_D}.\mathcal{LR} \gets $ \textproc{leastSquares}(\textproc{keys}($\boldsymbol{P}_{\mathsf{D}}$), $\tilde{[\mathsf{N_D}.\Omega]}$)}
		\State {$\mathsf{N_D}.\mathcal{LR}.a \gets \mathsf{N_D}.\mathcal{LR}.a \times r$, $\mathsf{N_D}.\mathcal{LR}.b \gets \mathsf{N_D}.\mathcal{LR}.b \times r$}
		\State {\textproc{LocalOpt}($\mathsf{N_D}, \boldsymbol{P}_{\mathsf{D}}$)}
		\EndIf
		\State {$\mathsf{N_D}.\kappa  = \frac{\mathsf{N_D}.\Delta}{\mathsf{N_D}.\Omega}$}
		\State {\textbf{return} $\mathtt{notExist}$}
		\EndFunction
	\end{algorithmic}
\end{algorithm}

However, many new nodes for conflicting keys may increase the depth of leaf nodes wildly. Thus, it is necessary to adjust the layout of some leaf nodes when the search performance degrades. 
We observe the relationship among $\mathsf{N_D}.\Delta$, $\mathsf{N_D}.\Omega$  and $\mathsf{N_D}.\kappa$, and uses a flexible strategy to decide if a leaf node $\mathsf{N_D}$ should be adjusted. 
When inserting a pair $p$ to $\mathsf{N_D}$, from this node, if the average number of entries need to be accessed to search for a pair (\emph{i.e.}, $\frac{\mathsf{N_D}.\Delta}{\mathsf{N_D}.\Omega}$) is larger than a pre-defined threshold (line~20), we think the insertions degrade the search performance. 
Thus, $\mathsf{N_D}$ is adjusted, which starts by collecting all pairs covered by $\mathsf{N_D}$ (line~21).
Then, the capacity of $\mathsf{N_D}.\boldsymbol{V}$ is enlarged and we train $\mathsf{N_D}$'s linear model accordingly such that conflicts will happen more rarely (lines~22-24).
Finally, we redistribute the pairs with the local optimization (line~25).

The pre-defined threshold is set to $\lambda \cdot \mathsf{N_D}.\kappa$ ($\lambda > 1$) where $\mathsf{N_D}.\kappa$

\noindent is the average number of accessed entries in search for a pair covered by $\mathsf{N_D}$ \textbf{after executing the last local optimization at $\mathsf{N_D}$}. In our experiments, $\lambda$ is set to 2. 
As a result, if the average cost per search \emph{w.r.t.} $\mathsf{N_D}$ doubles after a series of insertions, we deem some nodes under $\mathsf{N_D}$ become too deep and the performance of searching for relevant keys degrades dramatically. In this case, it is better to collect all pairs that $\mathsf{N_D}$ covers, retrain $\mathsf{N_D}.\mathcal{LR}$ and redistribute those pairs. 
Finally, the value of $\mathsf{N_D}.\kappa$ will be updated (line~26).

\begin{figure}[htb]
	\centering
	\includegraphics[width=0.44\textwidth, trim={5mm 6mm 5mm 6mm},clip]{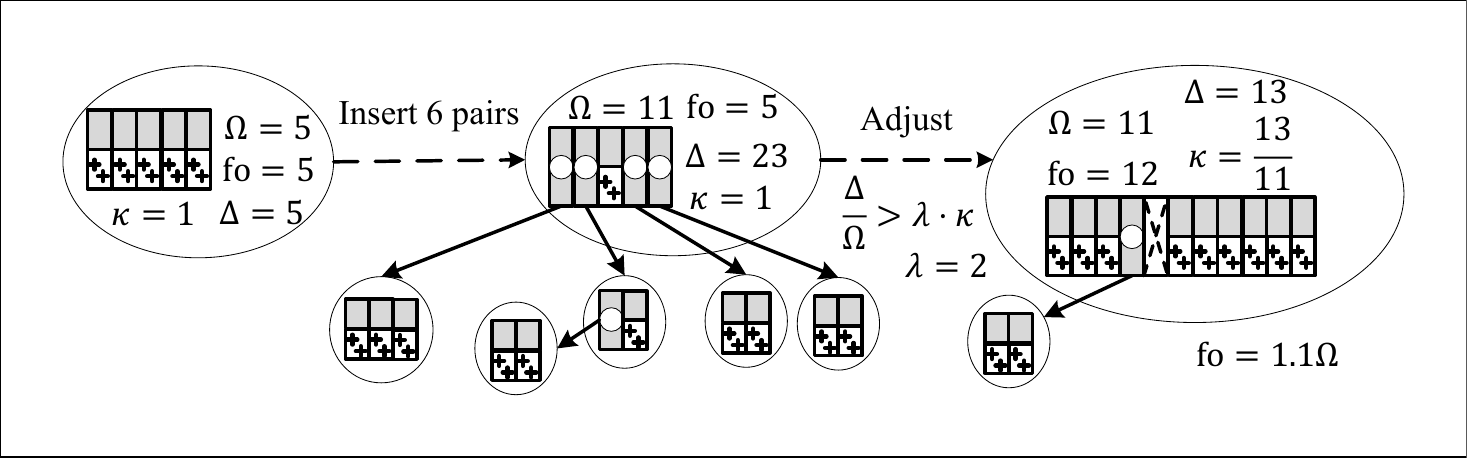}
	\caption{Adjusting a leaf node after insertions}
	\label{fig:node_adjust}
\end{figure}

When adjusting the leaf node $\mathsf{N_D}$, we also set $\mathsf{N_D}.\mathtt{fo}$ to be larger than $\mathsf{N_D}.\Omega$. The gap between them grows with more adjustments (line~22). 
A simple yet reasonable assumption is that the more adjustments, the more frequently relevant pairs are accessed. Also, more adjustments usually mean more conflicts. 
Thus, to reduce the number of conflicts at $\mathsf{N_D}$, we enlarge the capacity of $\mathsf{N_D}.\boldsymbol{V}$, making more slots for pairs. 
In our experiments, the enlarging ratio $\varphi(\mathsf{N_D}.\alpha) \triangleq \min(\eta + 0.1 \times \mathsf{N_D}.\alpha$, 4), where $\eta$ carries the same meaning with that in Algorithm~\ref{alg:LocalOpt}.
$\varphi(\cdot)$ can be any monotonically increasing function and its derivative should consider the memory usage. Our strategy of having redundancy in frequently adjusted nodes is similar to but more flexible than the usage of gapped array in ALEX~\cite{DBLP:conf/sigmod/DingMYWDLZCGKLK20}. Fig.~\ref{fig:node_adjust} gives an example of adjusting a leaf node.

\subsection{Deletions}
\label{subsec:deletion}

To delete a pair with the key $x$, we call \textproc{locateLeafNode} to find the highest leaf node $\mathsf{N_D}$ covering $x$ (line~1). Next, we use \textproc{deleteFromLeafNode} to delete the pair from $\mathsf{N_D}$ (line~2).
It recursively checks if there is a pair, covered by $\mathsf{N_D}$ or a leaf node underneath, having the key $x$ (lines~4-14). 
If the pair is found, we remove it by setting the corresponding slot in the leaf node's entry array to NULL (lines~5-6). 
Otherwise, \textproc{deleteFromLeafNode} simply returns False (lines~7-8).
After the removal, the values of $\mathsf{N_D}$'s some fields will change, like the number of pairs contained in $\mathsf{N_D}$ and the average cost of searching keys from $\mathsf{N_D}$. Thus, we update the values of $\mathsf{N_D}.\Delta$, $\mathsf{N_D}.\Omega$ and $\mathsf{N_D}.\kappa$ (lines~6, 10-12, 16).
If $\mathsf{N_D}$'s child leaf node $\mathsf{N'}$ contains only one pair $p''$ after the removal, we simply delete $\mathsf{N'}$ and replace the pointer to it with $p''$ in $\mathsf{N_D}.\boldsymbol{V}$ (lines~13-15).

\begin{algorithm}[htb]

\smaller
\caption{\textproc{Delete}($\mathsf{Root}, x$) \Comment{$x$ is the key to be deleted} }
\label{alg:delete}
\begin{algorithmic}[1]
	\smaller
	\State {$\mathsf{N_D} \gets$ \textproc{locateLeafNode}($\mathsf{Root}, x$) }
	\State {\textbf{return} \textproc{deleteFromLeafNode}($\mathsf{N_D}, x$)}
	
	\Function {deleteFromLeafNode} {$\mathsf{N_D}, x$}
	\State {$p' \gets \mathsf{N_D}.\boldsymbol{V}[pos]$, $pos \gets f_{\mathsf{D}} (x)$, $\mathtt{exist} \gets$ True}
	\If{$p'.key = x$ }
	\State {$\mathsf{N_D}.\boldsymbol{V}[pos] \gets$ NULL, $\mathsf{N_D}.\Delta$ -= 1} \Comment{delete $p'$ from $\mathsf{N_D}.\boldsymbol{V}$}
	\ElsIf {$p' =$ NULL}
	\State {$\mathtt{exist} \gets$ False}  \Comment{corresponding pair does not exsit}
	\ElsIf{$p'$ points to another leaf node $\mathsf{N}'$}
	\State {$\Delta' \gets \mathsf{N}'.\Delta$}
	\State {$\mathtt{exist} \gets$ \textproc{deleteFromLeafNode}($\mathsf{N}', x$)}
	\State {$\mathsf{N_D}.\Delta$ -= $1 + \Delta' - \mathsf{N'}.\Delta$}
	\If {$\mathsf{N}'.\Omega = 1$} \Comment{$\mathsf{N}'$ covers only one pair $p''$}
	\State {$\mathsf{N_D}.\boldsymbol{V}[pos] \gets$ the remaining one pair $p''$ contained in $\mathsf{N}'$ } 
	\State {$\mathsf{N_D}.\Delta$ -= 1 and delete $\mathsf{N}'$}
	\EndIf
	\EndIf
	\State {$\mathsf{N_D}.\Omega$ -= ($\mathtt{exist} =$ True ? 1 : 0), $\mathsf{N_D}.\kappa  = \frac{\mathsf{N_D}.\Delta}{\mathsf{N_D}.\Omega}$}
	\State {\textbf{return} $\mathtt{exist}$}
	\EndFunction
\end{algorithmic}
\end{algorithm}


\section{Experimental Studies}
\label{sec:experiments}

\modelName~and all competitors are implemented in C++~\cite{DILI_codes_url} and evaluated using a single thread on a Ubuntu server with a 96-core Xeon(R) Platinum 8163 CPU and 376 GB memory.
Due to space limit, we present more experimental results and analyses in an extended version~\cite{DBLP:journals/corr/abs-2304-08817}.

\subsection{Experimental Settings}

\textbf{Datasets.}~We use four real datasets from the SOSD benchmark~\cite{DBLP:journals/pvldb/MarcusKRSMK0K20} and one synthetic dataset.
\begin{itemize}[leftmargin=*]
	\item \textsf{FB}~\cite{FB_url} contains 200M Facebook user ids.
	\item \textsf{WikiTS}~\cite{WikiTS_url} contains 200M unique request timestamps (in integers) of log entries of the Wikipedia web-site.
	\item \textsf{OSM}~\cite{OSM_url} contains 800M ids of OpenStreetMap cells.
	\item \textsf{Books}~\cite{Books_url} contains 800M ids of books in Amazon.
	\item \textsf{Logn} contains 200M unique values sampled from a heavy-tail log-normal distribution with $\mu = 0$ and $\sigma = 1$. 
\end{itemize}
For each key, we associate it with a random integer number and pack them as a simulated record. The records are stored in an data array. 
For each record, its key and address together form a pair. For the pairs for index's bulk loading, we sort them according to their keys and feed them to index's bulk loading algorithm.

\noindent\textbf{Competitors.}~We compare~\modelName~with the following methods:
\begin{itemize}[leftmargin=*]
	\item \textbf{BinS} does a binary search over the whole sorted key set to find the position of the given search key.
	\item \textbf{B+Tree}~\cite{DBLP:journals/csur/Comer79}: We use a production quality B+Tree implementation stx::btree for comparison~\cite{btree_url}.
	\item \textbf{MassTree}~\cite{DBLP:conf/eurosys/MaoKM12} is a variant of B-Tree which improves cache-awareness by employing a trie-like~\cite{DBLP:journals/vldb/AskitisZ09} structure.
	\item \textbf{RMI}~\cite{DBLP:conf/sigmod/KraskaBCDP18} is built through linear stages and cubic stages. 
	\item \textbf{ALEX}~\cite{DBLP:conf/sigmod/DingMYWDLZCGKLK20} is an in-memory learned index which partition keys into leaf nodes in a relatively static way~\cite{ALEX_url}.
	\item \textbf{RS} (RadixSpline)~\cite{DBLP:conf/sigmod/KipfMRSKK020} uses a linear spline to approximate the CDF of the data and a radix table to index spline points. 
	\item \textbf{PGM} (PGM-index)~\cite{DBLP:journals/pvldb/FerraginaV20} contains  multiple levels, each representing an error-bounded piece-wise linear regression~\cite{PGM_url}.
	\item \textbf{LIPP}~\cite{DBLP:journals/pvldb/WuZCCWX21} can be seen as a special RMI. Its root node uses a linear regression model with the range of $[0, N)$, where $N$ is the dataset cardinality.
	At lower levels, LIPP recursively uses linear regression models to partition search keys until each key's position is accurately predicted.
	LIPP aims to predict as many keys' position as possible with only one model~\cite{LIPP_url}.
\end{itemize}
For RMI and RS, we adopt the implementations in SOSD~\cite{DBLP:journals/pvldb/MarcusKRSMK0K20,SOSD_url}.

Table~\ref{tab:model_properties} summarizes the properties of all indexes.
The better performance is indicated in bold.

\begin{smaller}
	\begin{table}[htb]
		\smaller
		\centering
		\caption{Properties of different methods}
		\label{tab:model_properties}
		\begin{tabular}{l|c|c|c|c|c}
			\hline 
			\textbf{Method} & \tabincell{c}{Support\\update} & \tabincell{c}{Consider data\\distribution} & \tabincell{c}{Extra local\\search} &  \tabincell{c}{Tree\\height} & \tabincell{c}{Memory\\cost}     \\ \hline
			B+Tree & $\boldsymbol{\checkmark}$ & $\times$ & $\checkmark$ & medium & medium \\ \hline
			RMI & $\times$ & $\times$ & $\checkmark$ & \textbf{low} & \textbf{small}  \\ \hline
			RS & $\times$ & $\boldsymbol{\checkmark}$ & $\checkmark$ & \textbf{low} & \textbf{small} \\ \hline
			PGM & $\boldsymbol{\checkmark}$ & $\times$ & $\checkmark$ & high & medium  \\ \hline
			Masstree & $\boldsymbol{\checkmark}$ & $\times$ & $\checkmark$ & medium &medium  \\ \hline
			ALEX & $\boldsymbol{\checkmark}$ & $\boldsymbol{\checkmark}$ & $\checkmark$ & medium & medium\\ \hline
			LPP & $\boldsymbol{\checkmark}$ & $\times$ & $\boldsymbol{\times}$ & medium  & large\\ \hline
			\modelName & $\boldsymbol{\checkmark}$ & $\boldsymbol{\checkmark}$  & $\boldsymbol{\times}$ & \textbf{low} & medium \\ \hline
		\end{tabular}
	\end{table}
\end{smaller}

\noindent\textbf{Evaluation Metrics.} We use two performance metrics: \textbf{Lookup} is the average lookup time per query, including the time spent in the index and in finding the records in the data array. \textbf{Throughtput} is the number of operations, including query, insertion and deletion, that of a method can handle per second.

\noindent\textbf{Parameter Settings.}~
Table~\ref{tab:exp_para_setting} lists the parameter settings for 
B+Tree and ALEX. They are built with bulk loading for better lookup and throughput performance.
For RMI and RS, we follow~\cite{DBLP:journals/pvldb/MarcusKRSMK0K20} to use two settings with the largest (L) and smallest (S) memory costs.

In our machine, an LL-cache line is of 64 bytes and fetching a cache line from the memory costs 130 CPU cycles at worst~\cite{corporporation2018intel,fog2018lists,intel_guide}. A~\modelName~(internal or leaf) node can be held in a single cache line. 
Therefore, we set $\theta_{\mathsf{N}} = \theta_{\mathsf{C}} = 130$. Executing a linear function as well as type casting cost about $\eta = 25$ cycles. Moreover, $\mu_{L} = 5$ and $\mu_{E} = 17$ cycles are spent on executing operations except accessing pairs in linear search and exponential search, respectively. 
The decaying rate $\rho$ in Eq.~\ref{Eq:bunode_lookup_cost}, enlarging ratio $\eta$ in Algorithm~\ref{alg:LocalOpt} and maximum fanout $\omega$ in Algorithm~\ref{alg:greedy_merge} are set to 0.2, 2 and 4,096, respectively.

\begin{smaller}
	\begin{table}[htb]
		\smaller
		\centering
		\caption{Parameter settings in experiments}\label{tab:exp_para_setting}
		\begin{tabular}{c|l|l}
			\hline 
			\textbf{Param} & \textbf{Description} & \textbf{Setting} \\ \hline
			$\Omega$ & Node fanout of a B+Tree & 16, 32, 64, 128, 256, 512\\ \hline
			$\Gamma$ & Max node size of ALEX & 16KB, 64KB, 1MB, 16MB, 64MB\\ \hline
		\end{tabular}
	\end{table}
\end{smaller}

\subsection{Overall Query Performance}
\label{subsec:overall_results}

For each dataset, we build all indexes using the whole dataset $\boldsymbol{P}$ and randomly select 100M keys in \textproc{keys}($\boldsymbol{P}$) to form point queries. All competitors are built with their preferred parameter settings.
Table~\ref{tab:comparison} reports on the overall performance results of all methods on point queries. 
To investigate the effect of the local optimization, we also include a DILI variant~\ModelWOLO~that applies no local optimization in its leaf nodes but tightly arranges pairs in the entry arrays. The search via~\ModelWOLO~simply follows Algorithm \ref{alg:search}.
We choose the LIPP as the fixed reference point as it is the best among all competitors. 
The color-encoding indicates how much faster or slower a model is against the reference point.

\begin{smaller}
	\begin{table}[ht]
		\smaller
		\centering
		\caption{Lookup time (ns) of all methods after bulk loading}
		\label{tab:comparison}
		\arrayrulecolor{black}
		\begin{tabular}{ L{1.2cm} V{3} L{1.1cm} V{3} r V{3} r V{3} r  V{3} r V{3} r}
			\specialrule{2pt}{0pt}{0pt}
			Model & Config & \multicolumn{1}{c V{3}}{FB} &  \multicolumn{1}{c V{3}}{WikiTS} &  \multicolumn{1}{c V{3} }{OSM} & \multicolumn{1}{c V{3} }{Books} & \multicolumn{1}{c }{Logn}  \\ \specialrule{2pt}{0pt}{0pt}
			\textbf{BinS} &  &
			\cellcolor{red3} 819 &
			\cellcolor{red3} 822 &
			\cellcolor{red3}839 &
			\cellcolor{red3}844 &
			\cellcolor{red3} 817  \\ \specialrule{1.5pt}{0pt}{0pt}
			
			\multirow{6}{*} {\textbf{B+Tree}} &
			$\Omega$=16 &
			\cellcolor{red3}629  &
			\cellcolor{red3}633 &
			\cellcolor{red3}578 &
			\cellcolor{red3}584 &
			\cellcolor{red2}624  \\
			\noalign{\makeatletter
				\global\let\CT@drsc@old\CT@drsc@
				\global\let\CT@drsc@\relax}
			\hhline{~------}
			\noalign{\makeatletter
				\global\let\CT@drsc\CT@drsc@old}
			
			& $\Omega$=32 &
			\cellcolor{red3}620 &
			\cellcolor{red3}616  &
			\cellcolor{red3}589 &
			\cellcolor{red3}611 &
			\cellcolor{red3}629  \\ \hhline{*{1}{|~}*{6}{|-}|}
			& $\Omega$=64 &
			\cellcolor{red3}658 &
			\cellcolor{red3}649 &
			\cellcolor{red3}641 &
			\cellcolor{red3}651 &
			\cellcolor{red3}653  \\
			\hhline{*{1}{|~}*{6}{|-}|}
			& $\Omega$=128 &
			\cellcolor{red3}722 &
			\cellcolor{red3}719 &
			\cellcolor{red3}693 &
			\cellcolor{red3}699 &
			\cellcolor{red3}725 \\
			\hhline{*{1}{|~}*{6}{|-}|}
			& $\Omega$=256 &\
			\cellcolor{red3}794 &
			\cellcolor{red3}790 &
			\cellcolor{red3}776 &
			\cellcolor{red3}775 &
			\cellcolor{red3}790 \\\hhline{*{1}{|~}*{6}{|-}|}
			& $\Omega$=512 &
			\cellcolor{red3}995&
			\cellcolor{red3}980 &
			\cellcolor{red3}979 &
			\cellcolor{red3}982 &
			\cellcolor{red3}984 \\ \specialrule{1.5pt}{0pt}{0pt}
			
			\multirow{5}{*} {\textbf{ALEX}}
			& $\Gamma$=16KB  &
			\cellcolor{red3}655 &
			\cellcolor{red3}580 &
			\cellcolor{red3}544 &
			\cellcolor{red3}509 &
			\cellcolor{red3}463  \\
			\noalign{\makeatletter
				\global\let\CT@drsc@old\CT@drsc@
				\global\let\CT@drsc@\relax}
			\hhline{~------}
			\noalign{\makeatletter
				\global\let\CT@drsc\CT@drsc@old} \
			& $\Gamma$=64KB   &
			\cellcolor{red3}573 &
			\cellcolor{red3}465 &
			\cellcolor{red2}419 &
			\cellcolor{red2}382 &
			\cellcolor{red3}398  \\
			\hhline{*{1}{|~}*{6}{|-}|}
			& $\Gamma$=1MB   &
			\cellcolor{red2}490 &
			\cellcolor{red1}248 &
			\cellcolor{red2}281 &
			\cellcolor{red2}274 &
			\cellcolor{red2}259  \\
			\hhline{*{1}{|~}*{6}{|-}|}
			& \cellcolor{gray}$\Gamma$=16MB  &
			\cellcolor{red2}476 &
			\cellcolor{red1}236 &
			\cellcolor{red1}223 &
			\cellcolor{red1}221 &
			\cellcolor{green1}170 \\
			\hhline{*{1}{|~}*{6}{|-}|}
			& $\Gamma$=64MB  &
			\cellcolor{red2}462 &
			\cellcolor{red1}252 &
			\cellcolor{red1}234 &
			\cellcolor{red1}203 &
			\cellcolor{green1}161 \\  \specialrule{1.5pt}{0pt}{0pt}
			
			\multirow{2}{*} {\textbf{RMI}}
			& (S)  &
			\cellcolor{red3}833 &
			\cellcolor{red3}806 &
			\cellcolor{red3}1255 &
			\cellcolor{red3}540 &
			\cellcolor{red3}907 \\
			\noalign{\makeatletter
				\global\let\CT@drsc@old\CT@drsc@
				\global\let\CT@drsc@\relax}
			\hhline{~------|}
			\noalign{\makeatletter
				\global\let\CT@drsc\CT@drsc@old}
			& (L)  &
			\cellcolor{red1}215 &
			\cellcolor{red1}175 &
			\cellcolor{red1}166&
			\cellcolor{red1}221 &
			\cellcolor{red1}208 \\ \specialrule{1.5pt}{0pt}{0pt}
			\multirow{2}{*} {\textbf{RS}}
			& (S)  &
			\cellcolor{red2} 398 &
			\cellcolor{red3} 313 &
			\cellcolor{red3}358 &
			\cellcolor{red3}355 &
			\cellcolor{green1}172 \\
			\hhline{*{1}{|~}*{6}{|-}|}
			\noalign{\makeatletter
				\global\let\CT@drsc@old\CT@drsc@
				\global\let\CT@drsc@\relax}
			\hhline{~----|}
			\noalign{\makeatletter
				\global\let\CT@drsc\CT@drsc@old}
			& (L)  &
			\cellcolor{red2}305 &
			\cellcolor{red1}264 &
			\cellcolor{red1}218 &
			\cellcolor{red1}210 &
			\cellcolor{green2}132 \\ \specialrule{1.5pt}{0pt}{0pt}
			\textbf{MassTree} &  &
			\cellcolor{red3} 1245&
			\cellcolor{red3} 1238 &
			\cellcolor{red3} 1500 &
			\cellcolor{red3} 1492&
			\cellcolor{red3} 1220 \\ \specialrule{1.5pt}{0pt}{0pt}
			\textbf{PGM} &  &
			\cellcolor{red3} 483 &
			\cellcolor{red3} 468 &
			\cellcolor{red3} 474 &
			\cellcolor{red3} 457 &
			\cellcolor{red3} 453  \\ \specialrule{1.5pt}{0pt}{0pt}
			
			\textbf{LIPP} &  &
			\cellcolor{grey} 197 &
			\cellcolor{grey} 152 &
			\cellcolor{grey} 178 &
			\cellcolor{grey} 182 &
			\cellcolor{grey} 173  \\ \specialrule{1.5pt}{0pt}{0pt}
			\textbf{\ModelWOLO} &  &
			\cellcolor{red1} 240 &
			\cellcolor{red1} 168 &
			\cellcolor{red1} 192&
			\cellcolor{red1} 208 &
			\cellcolor{green2} 142  \\ \specialrule{1.5pt}{0pt}{0pt}
			\textbf{\modelName} &  &
			\cellcolor{green4} \textbf{150} &
			\cellcolor{green4} \textbf{139} &
			\cellcolor{green4} \textbf{126} &
			\cellcolor{green4} \textbf{153} &
			\cellcolor{green4} \textbf{116} \\ \Xcline{1-7}{2pt}
		\end{tabular}
	\end{table} 
\end{smaller}

\modelName~has clear advantages over other state-of-the-art methods. 
Compared to LIPP,~\modelName~saves about 9\% to 34\% lookup time. The design of~\modelName's bulk loading algorithm make the keys in~\modelName's leaf nodes almost linearly distributed and the linear regression models well describe these distributions. Thus, conflicts happens more rarely in~\modelName. The traversal path of~\modelName~is shorter than that of LIPP and other competitors, which results in~\modelName~has better performance.
Compared to ALEX and PGM, besides the shorter traversal path, ~\modelName~is able to avoid the search inside the leaf nodes and have clearer advantages.  
BinS, MassTree and all variants of B+Tree even needs to take 4--10 times of lookup time to search for a key on average. 
Also, \modelName~clearly outperforms RMI and RS on processing point queries. And it is noteworthy that RMI and RS do not support updates.
These results illustrate~\modelName~achieves large lookup superiority over other alternatives.
In addition,~\modelName~consumes less lookup time than~\ModelWOLO~over all the five datasets. This verifies the effectiveness of the local optimization in~\modelName's leaf nodes.

As B+Tree with $\Omega = 32$, ALEX with $\Gamma = 16$MB and the large RMI and RS perform best among their variants, we will choose them as representatives and omit the evaluation of the other variants in the following sections. The parameter settings of these methods will also be omitted when we refer to them. Also, to save space, we will omit the comparable results on the datasets OSM and Books. The results on both datasets are similar to that on other datasets. 

\noindent\underline{\textbf{Cache Misses.}}~\modelName's advantage is partly due to that the design of~\modelName's structure makes~\modelName~triggers fewer cache misses. A single LL-cache miss incurs 50-200 additional cycles~\cite{corporporation2018intel,intel_guide,DBLP:conf/sigmod/KraskaBCDP18}. In contrast, register operations like addition and multiplication cost 1-3 cycles only. Avoiding cache misses clearly speeds up query processing for \modelName. Table~\ref{tab:cache_misses} reports the average number of LL-cache misses for all methods. 
In particular, compared with ALEX and LIPP,~\modelName~avoids up to 9.7 and 3.5 LL-cache misses per query.

\begin{smaller}
	\begin{table}[htb]
		\smaller
		\centering
		\caption{\#LL-cache misses of methods per point query}
		\label{tab:cache_misses}
		\begin{tabular}{l | c | c | c | c | c | c | c | c}
			\hline 
			Dataset &  B+Tree & RMI & RS & PGM & MassTree & ALEX & LIPP & \modelName \\ \hline
			\textsf{FB}  & \cellcolor{red2}10.27 & \cellcolor{green2}5.25 & \cellcolor{red1}8.43  & \cellcolor{red2} 10.73& \cellcolor{red2} 9.84& \cellcolor{red3}14.91 & \cellcolor{grey}7.94 & \cellcolor{green4}\textbf{4.88} \\ \hline
			\textsf{WikiTS}  & \cellcolor{red3}10.51 & \cellcolor{green4}\textbf{4.60} & \cellcolor{green2}5.68 & \cellcolor{red3} 11.56 & \cellcolor{red3} 9.24& \cellcolor{red2}7.36 & \cellcolor{grey}5.86 & \cellcolor{green4}4.78 \\ \hline
			\textsf{Logn} & \cellcolor{red3}10.19 & \cellcolor{green2}5.28 & \cellcolor{green4}\textbf{3.22} & \cellcolor{red3} 9.88 & \cellcolor{red3} 9.48 & \cellcolor{green2}4.47 & \cellcolor{grey} 7.17 & \cellcolor{green3}3.80 \\ \hline
			\textsf{OSM} & \cellcolor{red3}10.47 &  \cellcolor{green3}\textbf{3.89} &  \cellcolor{green2}4.50 &  \cellcolor{red2}7.42 &  \cellcolor{red3}12.85 & \cellcolor{green2}5.86 & \cellcolor{grey}7.13 & \cellcolor{green4}4.08\\ \hline
			\textsf{Books} & \cellcolor{red3}10.46 &  \cellcolor{green2}6.02 &  \cellcolor{green2}4.74 &  \cellcolor{green2}7.38 &  \cellcolor{red3}13.02 & \cellcolor{green3}\textbf{4.27} & \cellcolor{grey}7.81 & \cellcolor{green4}4.31\\ \hline	\end{tabular}
	\end{table}
\end{smaller}

\noindent\underline{\textbf{Offline Construction Time.}} 
On a 100M dataset, the bulk loading of B+Tree, ALEX, LIPP and~\modelName~takes less than 1, 2, 1 and 6 minutes, respectively. Each construction time grows almost linearly with the increase of the data size. 
The most time-consuming step in~\modelName's construction is the greedy merging algorithm to get the BU nodes at the bottom layer. A direct yet effective approach to make this step more efficient is sampling.
When a piece $I_{u}^{k}$ (in Algorithm~\ref{alg:greedy_merge}) covers many keys, we could randomly or selectively sample part of the keys, \emph{e.g.,} select one key out of two, to get the linear regression model and calculate the cost. The sampling strategy makes little influence on the whole BU-tree node layout or the performance of the generated DILI. However, it will make the the construction time of the BU-tree and~\modelName~decrease by over 1 minute.
As the bulk loading is one-time,~\modelName's a few more minutes highly pay off.

\begin{smaller}
	\begin{table}[htb]
		\smaller
		\centering
		\caption{The statistics of~\modelName}
		\label{tab:model_stats}
		\begin{tabular}{l | c | c | c | c}
			\hline 
			Dataset &  \tabincell{c}{Minimum\\height} & \tabincell{c}{Maximum\\height} & \tabincell{c}{Average\\height}  & \tabincell{c}{\# of conflicts\\ per 1K keys}  \\ \hline
			\textsf{FB} & 3 & 8 & 3.45 & 227.1 \\ \hline
			\textsf{WikiTS}  & 3 & 6 & 3.09 & 44.4 \\ \hline
			\textsf{Logn} & 3 & 4 & 3.01 & 1.2 \\ \hline
			\textsf{OSM} & 3 & 9 & 3.26 & 117.7 \\ \hline
			\textsf{Books} & 3 & 8 & 3.44 & 220.4\\ \hline	\end{tabular}
	\end{table}
\end{smaller}

\noindent\underline{\textbf{Analysis of~\modelName's Construction.}} Table~\ref{tab:model_stats} shows~\modelName's minimum/ maximum/average heights and the number of conflicts in~\modelName's construction for different datasets. Apparently,~\modelName~has a shallow structure. The slot assignments for most pairs cause no conflicts.
The average heights of the~\modelName s built on the Logn and WikiTS dataset are smaller than that of others. The reason is that the keys in both datasets are more linearly or piecewise linearly distributed. Thus, the linear regression models in the leaf nodes are able to make more accurate predictions and thus result in less conflicts.

\begin{figure}[htb]
	\centering
	\includegraphics[width=0.3\textwidth, trim={15mm 135mm 150mm 0mm},clip]{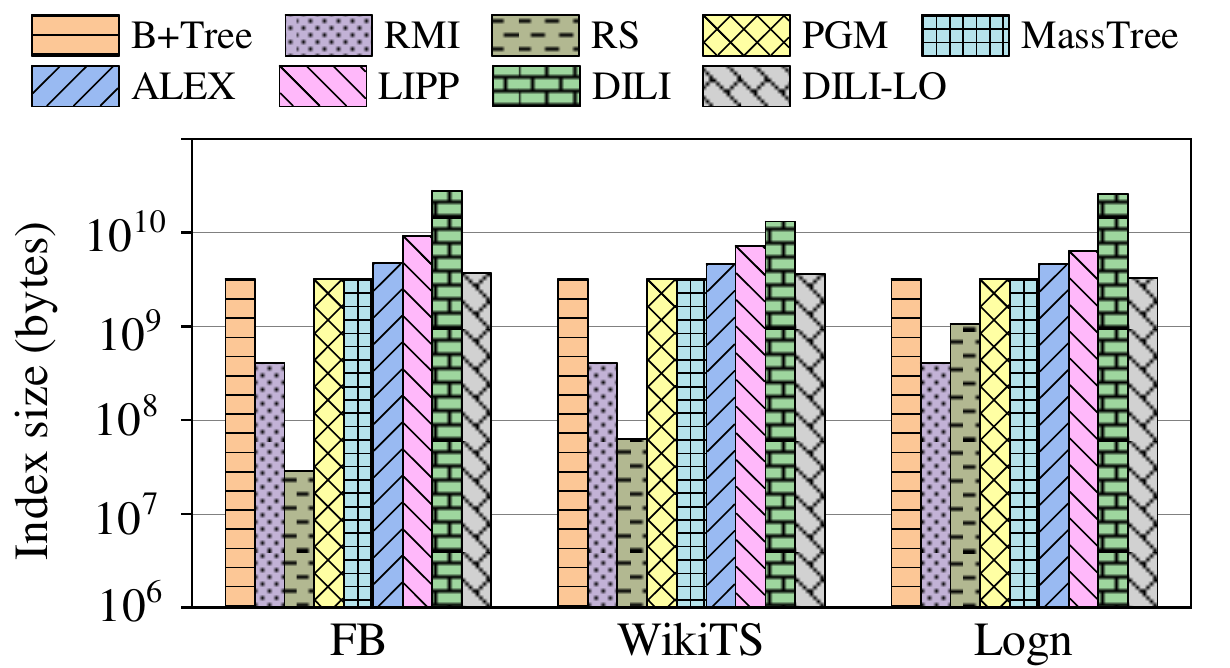}
	\caption{Index sizes}
	\label{fig:size}
\end{figure}

\noindent\underline{\textbf{Index Size.}} Fig.~\ref{fig:size} displays the memory cost of different methods. 
RMI and RS consume the least memory. However, they do not need to store pairs in their structures and do not support data updates.~\modelName~consumes more memory than B+Tree, PGM and ALEX due to the local optimization in the leaf nodes. A conflict will result in a new leaf node creation and an empty slot in the entry array. 
Nevertheless, our design strikes a trade-off between the memory cost and the query efficiency.  
Considering modern computers usually have huge memory, it is acceptable to improve the efficiency at the expense of some memory. Although LIPP also tries to strike a memory-efficiency trade-off and adopts a similar strategy for conflicts, its node layout is not so optimized as ours. Thus, LIPP results in more conflicts and memory costs.  Its memory cost is at least one order of magnitude larger than others.
On the other hand, after disabling the local optimization, the memory cost of the variant~\ModelWOLO~becomes comparable with B+Tree. Meanwhile, the query performance does not degrade much. 

\subsection{Performance on Different Workloads}
\label{subsec:exp_workloads}

\begin{figure*}[!htb]
	\centering
	\includegraphics[width=0.96\textwidth, trim={0mm 6mm 0mm 0mm},clip]{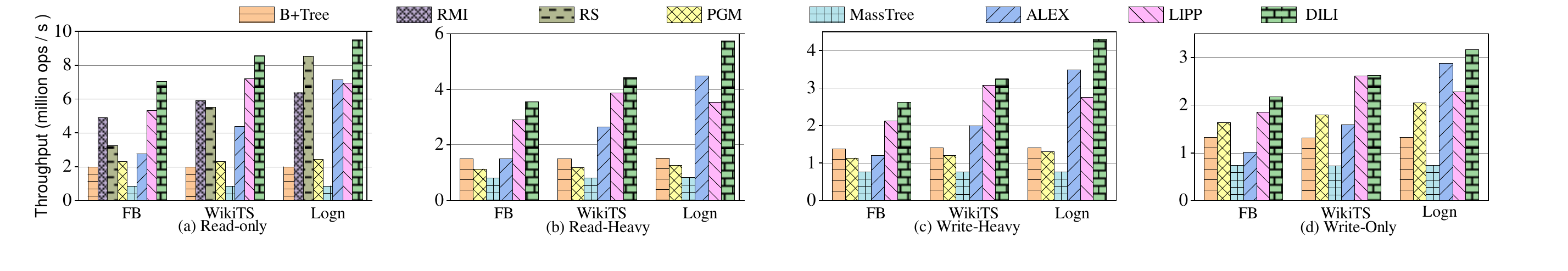}
	\caption{\modelName~vs. State-of-the-art methods: throughput comparisons on four workloads}
	\label{fig:throughputs}
\end{figure*}

We conduct experiments to compare~\modelName~and the alternatives on different types of workloads: (1) The Read-only workload contains 100M point queries. (2) The Read-Heavy workload contains 50M insertions and 100M point queries. (3) The Write-Heavy workload contains 100M insertions and 50M point queries. (4) The Write-only workload contains 100M insertions. 
In each dataset $\boldsymbol{P}$, we randomly select 50\% of the pairs as the initial dataset $\boldsymbol{P}_{0}$.  The other 50\% of $\boldsymbol{P}$ is named $\boldsymbol{P}_{1}$.
All workloads are tested on an index with bulk loading of $\boldsymbol{P}_{0}$. Besides, the query keys are randomly selected from the $\textproc{keys}(\boldsymbol{P})$, and the pairs to be inserted are randomly chosen from $\boldsymbol{P}_{1}$. 
Each workload is a random mix of queries and insertions.
We run the workloads on different indexes for five times and obtain their average throughput.
As RMI and RS do not support updates, they are excluded from the experiments involving insertions. The experimental results are shown in Fig.~\ref{fig:throughputs}. Overall,~\modelName~achieves the highest throughput on all workloads.

For the read-only workloads, compared to the others,~\modelName~achieves shorter average search path. In particular,~\modelName~accesses only 0.2-1 node per point query on average. This indicates that~\modelName~utilizes the data distribution well and thus the learned models in its leaf nodes incur few conflicts.
The alternatives need longer search paths queries and extra steps to carry out local search.
RMI and RS achieves comparably long search paths with~\modelName. However, the effort of correcting their prediction results in lower throughput.

When more insertions are in the workloads, we see that~\modelName~still outperforms others though its performance also degrades. 
The reason is that an insertion not only includes searching for a key but also writing a pair to an entry array. Also, new node creations are required to process conflicts. Moreover, adjustments occasionally happen to bound~\modelName's height.
Even though insertions on~\modelName~requires more time than queries,~\modelName~is still able to deal with index structure change well and more efficient at insertions than others.
PGM performs worst in these workloads as it needs $O(\log N)$ trees to support insertions and each query will search in all these trees.
Compared to B+Tree and ALEX,~\modelName~can avoid element shifting. Also, the new node creation in~\modelName~is light-weight.
In addition, compared to LIPP,~\modelName~has shorter traversal path for insertions. Our strategy of setting more slot redundancy for leaf nodes more frequently accessed also avoid unnecessary node adjustments.

\subsection{Effect of Many Deletions}
\label{subsec:exp_delete}

\begin{figure}[htb]
	\centering
	\includegraphics[width=0.43\textwidth, trim={0mm 100mm 0mm 0mm},clip]{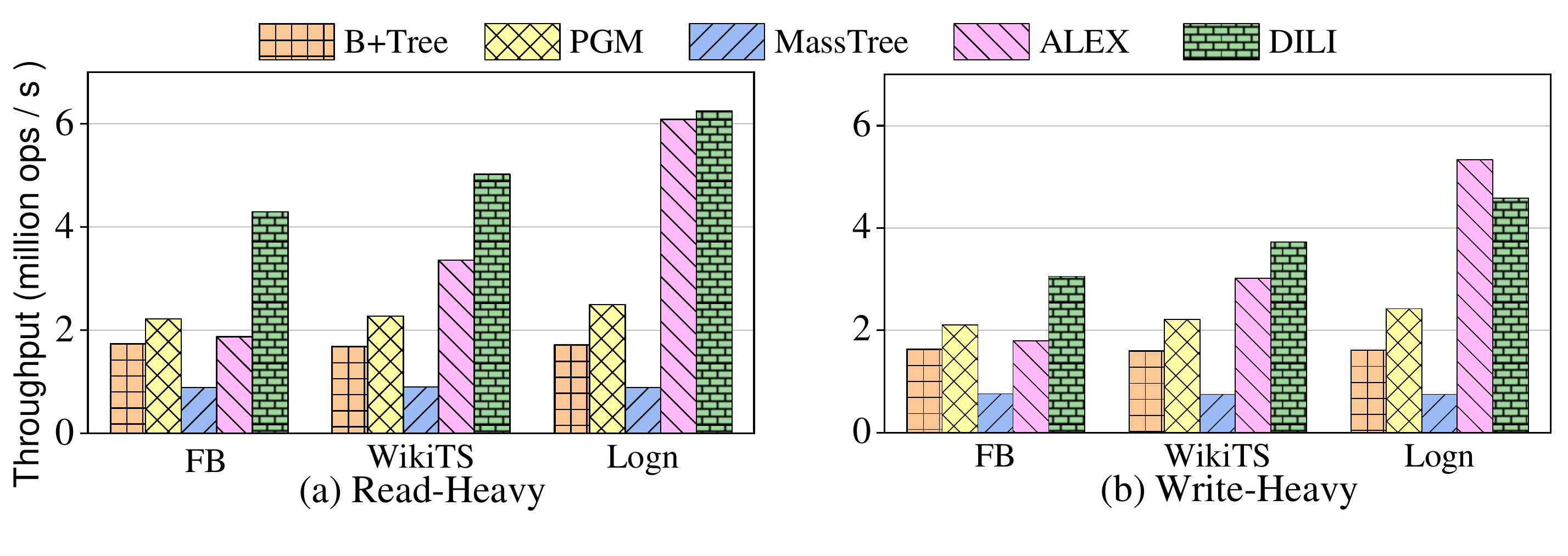}
	\caption{Performance after deletions}
	\label{fig:del_results}
\end{figure}

We also experimentally investigate the effect of deletions on~\modelName, B+Tree, PGM MassTree and ALEX. LIPP is excluded as it does not support deletions.
We first build each of them with bulk loading of the whole $\boldsymbol{P}$. Then we repeatedly delete/search for random keys from $\boldsymbol{P}$ via all methods and observe their changing throughput on three workloads: (1) Read-Heavy workload, which contains 100M lookups and 50M deletions; (2) Deletion-Heavy workload, which contains 100M deletions and 50M loopups.
Fig.~\ref{fig:del_results} shows the results.

Referring to Figure~\ref{fig:del_results}, on Read-Heavy workload,~\modelName~achieves up to 3.6$\times$, 2.3$\times$, 7.0$\times$ and 2.3$\times$ higher throughput than B+Tree, PGM, MassTree and ALEX, respectively. This illustrates that~\modelName~maintains high performance on queries with deletions happening.
On Deletion-Heavy workload, only ALEX performs a little better than DILI on Logn dataset. As ALEX almost adopts lazy deletion strategy, deleting a pair from ALEX almost equals searching for it. However, this strategy will cause its lookup time not decrease even through it index a small amount of data only.  Actually,~\modelName~performs much better than ALEX when the workload consists of more queries.


\section{Related Work}\label{sec:related_work}

\textbf{B-tree variants.} 
A B+Tree~\cite{DBLP:journals/csur/Comer79} is the most popular B-tree variant in which each internal node contains only keys, and the leaf nodes are chained with extra links.
Digital B-trees~\cite{DBLP:conf/vldb/Lomet81} allows a node to use two pages via a hashing-like technique.
The B-trie~\cite{DBLP:journals/vldb/AskitisZ09} combines B-Tree and trie~\cite{knuth1997art} to index strings stored in external memory.
MassTree~\cite{DBLP:conf/eurosys/MaoKM12} employs a trie-like concatenation of B-trees to improve cache-awareness in indexing key-value pairs. 
A BF-tree~\cite{DBLP:journals/pvldb/AthanassoulisA14} replaces B-Tree leaf nodes with bloom filters to substantially reduce the index size.
Unlike all B-tree variants, our~\modelName~stores models instead of pointers in the nodes for indexing purpose.

\noindent \textbf{Learned indexes for 1D keys.} 
The recursive model index (RMI)~\cite{DBLP:conf/sigmod/KraskaBCDP18}  uses staged models. An internal model directs a key search to one of its child models and a bottom-level model predicts an error-bounded position in the database.
RMI has inspired a number of learned indexes.
To reduce index memory footprint, a FITing-Tree~\cite{DBLP:conf/sigmod/GalakatosMBFK19} uses linear models to replace the leaf nodes of a B-Tree.
CARMI~\cite{DBLP:journals/pvldb/ZhangG22} applies data partitioning to RMI construction and supports data update.
NFL~\cite{DBLP:journals/pvldb/WuCYSKX22} uses a normalizing flow techniques~\cite{tabak2013family} to transform the key space for better approximation on the CDF.
PGM-index~\cite{DBLP:journals/pvldb/FerraginaV20} employs piecewise linear models to approximate the relationship between search keys and their positions in a database. 
Hermit~\cite{DBLP:conf/sigmod/WuYTSB19} creates a succinct tiered regression search tree (TRS-tree) which passes a search query to an existing index for correlated columns.
RadixSpline~\cite{DBLP:conf/sigmod/KipfMRSKK020} uses a set of spline functions as the learned index that can be built in a single pass over sorted data. 
ALEX~\cite{DBLP:conf/sigmod/DingMYWDLZCGKLK20} trains accurate linear regression models to split the key space, organizes all models also in a tree-like structure, and uses a gapped array for each leaf node. ALEX supports updates.
LIPP~\cite{DBLP:journals/pvldb/WuZCCWX21} uses kernelized linear functions as learned models that make perfect predictions. However, it does not make use of the information of data distribution.
SOSD~\cite{DBLP:journals/pvldb/MarcusKRSMK0K20} is a preliminary benchmark for 1D learned indexes.
FINEdex~\cite{DBLP:journals/pvldb/LiHJZ21} is a fine-grained learned index scheme, which constructs independent models with a flattened data structure to process concurrent requests.
APEX~\cite{DBLP:journals/pvldb/LuDLMW21} combines the recently released persistent memory optimization~\cite{Intel_PMem} and ALEX to support persistence and instant recovery.
The on-disk learned index prototype AirIndex~\cite{DBLP:conf/sigmod/Chockchowwat22} uses with a storage-aware auto-tuning method to minimize accesses to the external memory.
To validate the effectivenesss of the existing updatable learned indexes, Wongkham et al.~\cite{DBLP:journals/pvldb/WongkhamLLZLW22} conduct a comprehensive evaluation.

\noindent \textbf{Learned indexes for multidimensional data.} SageDB~\cite{DBLP:conf/cidr/KraskaABCKLMMN19} extends RMI to index multidimensional data in a transformed 1D space.
ZM-index~\cite{DBLP:conf/mdm/WangFX019} applies RMI to the Z-order curve~\cite{morton1966computer} to process spatial point and range queries.
ML-index~\cite{DBLP:conf/edbt/DavitkovaM020} applies RMI to iDistance~\cite{DBLP:journals/tods/JagadishOTYZ05} to support queries on multidimensional data.
Flood~\cite{DBLP:conf/sigmod/NathanDAK20} and Tsunami~\cite{DBLP:journals/corr/abs-2006-13282} are learned indexes for in-memory multidimensional data, whereas LISA~\cite{DBLP:conf/sigmod/Li0ZY020} and RSMI~\cite{DBLP:journals/pvldb/QiLJK20} are for disk-resident dynamic spatial data.
In contrast, our~\modelName~focuses on 1D data.


\section{Conclusion and Future Work}
\label{sec:conclusion}

In this work, we design for in-memory 1D keys a distribution-driven learned tree~\modelName. Its nodes use linear regression models to map keys to corresponding children or records. An internal node's key range is equally divided by all its children, endowing internal models with perfect accuracy for finding the leaf node covering a key.
We optimize~\modelName's node layout by a two-phase bulk loading approach. First, we create a bottom-up tree that balances the number of leaf nodes and tree height. Based on that, we determine for each~\modelName~internal node its best fanout and local model.
Also, we design algorithms for~\modelName~for data updates.
Extensive experimental results show that~\modelName~clearly expand the state of the art.

For future research, it is relevant to adapt~\modelName~to disk-resident data.
To this end, the cost model for the BU-Tree construction should consider the expected IOs, striking a trade-off between the IO cost and the computational overhead. Also, the local optimization should be disabled as it may create leaf nodes covering few keys.
\noindent Moreover, it is interesting to consider concurrent data updates with~\modelName.
Note that an insertion or deletion operation in~\modelName~involves only one leaf node. The node adjustment of~\modelName~is much simpler than the rebalance operation of the B+Tree. Theoretically, the lock-free and lock-crabbing~\cite{DBLP:journals/tods/Graefe10} approaches can be applied to~\modelName, in the same way as how they are applied to the B+Tree.

\begin{acks}
	Hua Lu’s work was partly supported by Independent Research Fund Denmark (No. 1032-00481B). 
\end{acks}

\balance

 \if\includeAppendix1
\newpage

\appendix
\nobalance
\section{Appendix}
\label{sec:appendix}
This appendix presents more experimental results and analyses that cannot be included in the conference version due to space limit.

\subsection{Index Scalability on Read-Only Workloadss}
\label{app:scalibility}
To further investigate how different methods perform in terms of scalability, we again run the read-only workloads, initializing each index with 50M, 100M, 150M and 200M keys, respectively. As the results on all datasets are similar, we only report the results on FB in Fig.~\ref{fig:scalibility_and_dist_shift} (a). As the number of indexed keys increases,~\modelName~maintains higher throughput than the alternatives. 
This indicates that~\modelName~is more adaptive to the size of data.

\begin{figure}[htb]
	\centering
	\vspace{-8pt}
	\includegraphics[width=0.5\textwidth, trim={15mm 116mm 10mm 0mm},clip]{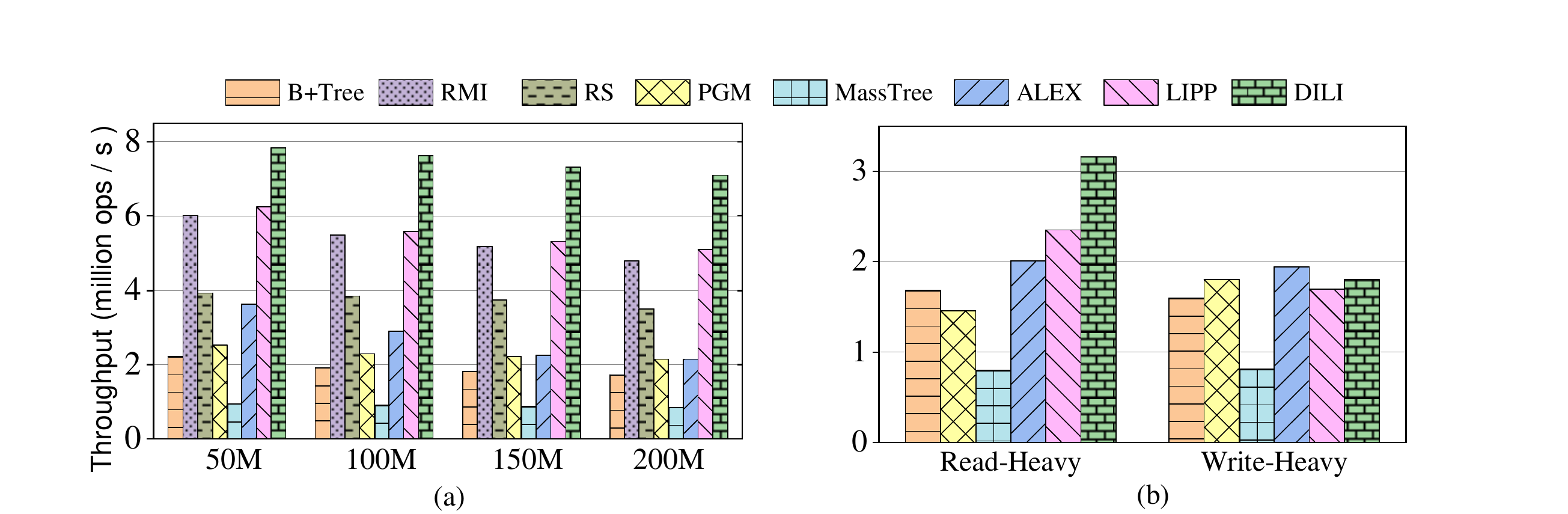}
	\vspace{-7pt}
	\caption{(a) Throughput on FB dataset with varying data cardinalities; (b) Performance with distribution shifting.}
	\label{fig:scalibility_and_dist_shift}
	\vspace{-2pt}
\end{figure}

\subsection{Effect of Distribution Shifting}
\label{app:diff_dist}

To investigate if~\modelName~still works well when keys from a different distribution are inserted, we design an experiment as follows.
Suppose $\boldsymbol{P}_{A}$ is another pair set whose keys are in a different distribution from $\boldsymbol{P}$.
First, we build~\modelName~and other indexes supporting inserts using their bulk loading algorithms on the whole $\boldsymbol{P}$.
Then, we repeatedly insert random keys from $\boldsymbol{P}_{A}$ via all indexes. Meanwhile, search operations are alternatively conducted.
We tested a representative combinations of $\boldsymbol{P}$ and $\boldsymbol{P}_{A}$ from different distributions:  \textsf{FB} and \textsf{Logn}.
We observe each method's throughput on Read-Heavy and Write-Heavy workloads. Fig.~\ref{fig:scalibility_and_dist_shift} (b) reports that~\modelName~performs a bit worse than ALEX on Write-Heavy workload. 
The reason behind is that~\modelName~is built to grasp the distribution characteristics of specific dataset. Thus, inserting pairs from different distribution will incur more conflicts and result in adjustments happen more frequently.
However, on Read-Heavy workload,~\modelName~still has clear advantages over other alternatives. Also, in real-life scenarios, queries are often more common than insertions. Thus,~\modelName~is supposed to have better performance than other state-of-the-art methods in practice.

\subsection{Effect of Skewed Writes}
\label{app:extreme_dist_shift}

Our next experiment investigates the performance of~\modelName~and other alternatives with skewed writes.
Suppose $\boldsymbol{P}$, whose size is 100M, is the pair set used in the bulk loading stage. The range of \textproc{keys}($\boldsymbol{P}$) is $[A, A + 10\delta)$ and $\boldsymbol{Q}$ is another pair set whose keys are in a different distribution from $\boldsymbol{P}$. We first compress the keys in \textproc{keys}($\boldsymbol{Q}$) by mapping them to the range $[A, A + \delta)$. 
Next, a new pair set $\boldsymbol{Q'}$ is generated by randomly selecting 100M distinct mapped keys and packing them with the original record pointers of $\boldsymbol{Q}$ together. We use a $\boldsymbol{Q}$ with 200M elements to generate $\boldsymbol{Q'}$.
~\modelName, B+Tree, ALEX and LIPP are built using their bulk loading algorithms on the $\boldsymbol{P}$.
Afterwards, random pairs from the skewed set $\boldsymbol{Q'}$ are repeatedly inserted  and then search operations are conducted via all indexes.
We test all possible combinations of $\boldsymbol{P}$ and $\boldsymbol{Q'}$ from three different distributions: FB, WikiTS and Logn.
We observe each method's throughput on read and write operations. Fig.~\ref{fig:skewed_insertions} reports the results of DILI and the competitors.
\begin{figure}[htb]
	\centering
	\includegraphics[width=0.45\textwidth, trim={35mm 30mm 108mm 18mm},clip]{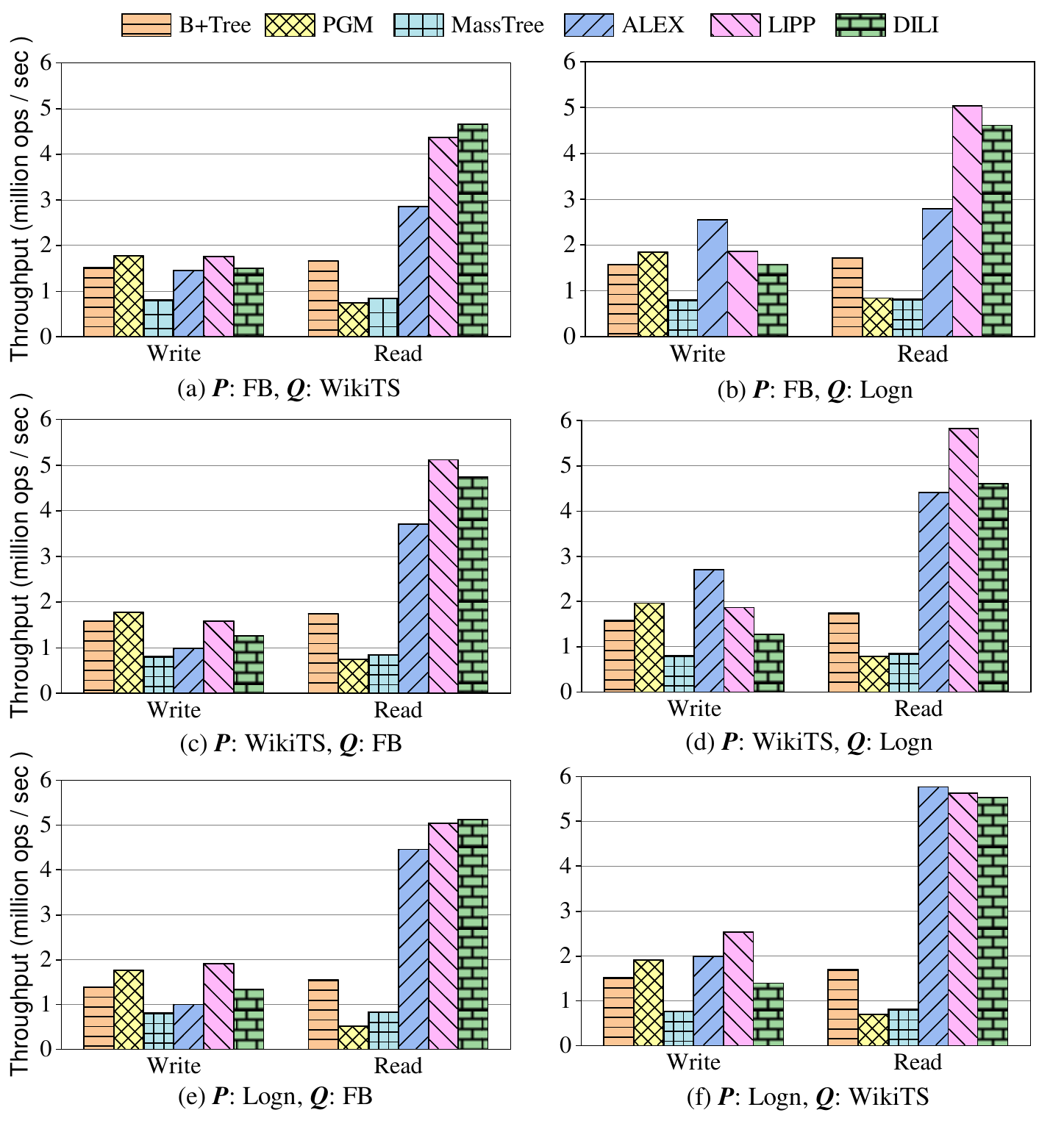}
	\vspace{-2pt}
	\caption{The effects of skewed insertions} 
	\label{fig:skewed_insertions}
	\vspace{-5pt}
\end{figure}

Referring to Fig.~\ref{fig:skewed_insertions}, DILI does not have clear advantages over other competitors as before. It is because that DILI is a distribution-driven learned index which has customized node structure for each dataset. When the distribution of the inserted keys is skewed, more conflicts and node adjustments tend to happen, which result in a higher tree. For example, after inserting 100M skewed keys from the WikiTS dataset to the DILI built with the FB dataset, the average height of the DILI is 4.22. In contrast, if the inserted keys are not out of the initial distribution, the height is 3.86. Nevertheless, in comparison with LIPP and ALEX, DILI still achieves comparable or even better results.

\subsection{Memory Cost Analysis in Write-heavy Workloads}
\label{app:memory_in_write_heavy}

\noindent We also conduct experiments to investigate the effect of insertions on the memory costs of~\modelName~and the alternatives. 
Similarly, we build all indexes with half of the dataset and use them to carry out 100M insertions where the inserted pairs are from the remaining half of the dataset. Table~\ref{tab:insert_effects} shows the memory cost comparisons among all indexes.
\begin{table}[htb]
	\centering
	\smaller
	\caption{The memory costs ($10^9$ bytes) of different indexes in write-heavy workloads}
	\label{tab:insert_effects}
	\setlength\tabcolsep{3pt}
	\arrayrulecolor{black}
	\begin{tabular}{ l | l | r | r | r | r | r | r}
		\hline
		Dataset & \tabincell{c}{Before/after\\ insertions}  & \tabincell{c}{B+Tree \\ $\Omega$=32} & MassTree & PGM & \tabincell{c}{ALEX \\ $\Gamma$=16MB} &  LIPP & \modelName \\ \hline
		
		\multirow{2}{*} {FB}
		& Before & 1.56 & 1.56 & 1.60 & 2.35 & 12.68 & 5.07\\ \cline{2-8}
		& After & 3.12 & 3.12 & 3.21 & 4.59 & 22.60 & 8.91\\ \hline
		\multirow{2}{*} {WikiTS}
		& Before & 1.56 & 1.56 & 1.60  & 2.31 & 10.11 & 3.95\\ \cline{2-8}
		& After & 3.12 & 3.12 & 3.20 & 4.31 & 21.22 & 5.79\\ \hline
		\multirow{2}{*} {Logn}
		& Before & 1.56 & 1.56 & 1.60 & 2.30 & 14.53 & 3.38\\ \cline{2-8}
		& After & 3.12 & 3.12 & 3.20 & 4.10 & 18.04 &3.67 \\ \hline
	\end{tabular}
\end{table}

As shown in Table~\ref{tab:insert_effects}, the memory costs of B+Tree, MassTree and PGM are smaller than that of ALEX,~\modelName~and LIPP.~\modelName~achieves comparable results with ALEX. 
After the same 100M insertions from the Logn dataset,~\modelName~uses less memory than ALEX. Compared to~\modelName, the memory costs of LIPP on all of the three datasets are much larger. This indicates that~\modelName~is able to avoid much more conflicts as well as the empty slots in the entry array.

\subsection{Range Query Performance}
\label{app:range_query_results}

The range query via~\modelName~is processed by a search for the lower bound key followed by a scan for the subsequent keys.
Fig.~\ref{fig:range_query}  reports the average response time of B+Tree, PGM, ALEX, LIPP, DILI and~\ModelWOLO~on short range queries. 
Following the settings in~\cite{DBLP:conf/sigmod/DingMYWDLZCGKLK20}, we use less than 100 keys in a range.
All methods are built with bulk loading on a full $\boldsymbol{P}$ on which 10M random range queries are issued.
The advantage of~\modelName~is less apparent than that in the point query performance comparison.
This is attributed to that the pairs are not densely stored in the entry arrays in~\modelName's leaf nodes and~\modelName~needs to distinguish between different entry types.
Nevertheless,~\modelName~achieves higher throughput than LIPP on all cases, and it is comparable to other competitors.
Also, the performance of~\modelName~could be improved by the variant~\ModelWOLO.
In this case, the leaf nodes' entry arrays only cover pairs. Accordingly, we only need to perform a sequential scan over the whole entry array, without handling other kinds of entry elements possible in~\modelName. Referring to Fig.~\ref{fig:range_query}, the average range query response time of~\ModelWOLO~is shorter than that of~\modelName.

\begin{figure}[htb]
	\centering
	\includegraphics[width=0.3\textwidth, trim={5mm 115mm 150mm 0mm},clip]{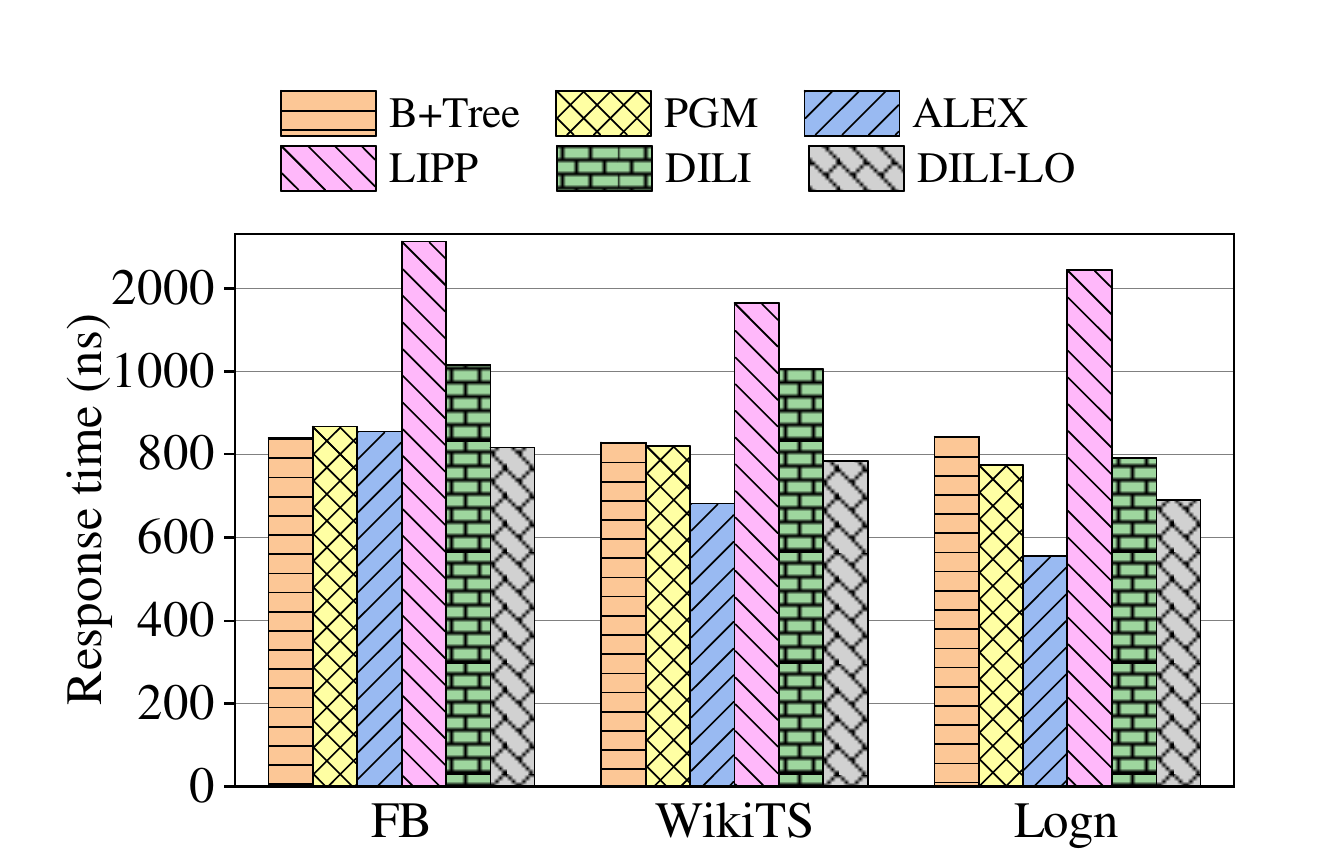}
	\caption{Range query results}
	\label{fig:range_query}
\end{figure}

\subsection{Hyperparameter Studies}
\label{subsec:exp_param}

To study the effects of hyperparameters, we build different~\modelName~ver-

\noindent sions accordingly and observe their performance on the FB dataset. The results on the other datasets are similar and thus omitted.

\noindent\underline{\textbf{Effects of $\omega$.}} $\omega$ is used to control the least number of nodes at~\modelName's each level. According to our observations, the generated~\modelName s with the value of $\omega$ varied from 1,024 to 8,192 have the same node layout. ~\modelName~tends to have a wide structure. Thus, as long as the value of $\omega$ is large enough, it slightly influences the performance.

\noindent\underline{\textbf{Effects of $\rho$.}} Table~\ref{tab:rho_effects} shows the lookup time and memory costs of~\modelName~with different values of $\rho$, the decaying rate of the impact of BU internal nodes at higher level on the layout of~\modelName's leaf nodes.
Apparently, the value of $\rho$ has little influence on~\modelName's overall structure and query performance. When the value of $\rho$ is set to around 0.1,~\modelName~performs the best.
\begin{smaller}
	\begin{table}[htb]
		\centering
		\smaller
		\caption{The effects of the hyperparameter $\rho$}
		\label{tab:rho_effects}
		\arrayrulecolor{black}
		\begin{tabular}{ l | r | r | r}
			\hline
			Param & \tabincell{c}{lookup \\ time (ns)} & \tabincell{c}{Memory cost \\ ($10^9$ bytes)} & \tabincell{c}{Average\\height} \\ \hline
			$\rho = 0.05$ & 154 & 9.327 & 3.441 \\ \hline
			$\rho = 0.1$ & 151 & 9.325 & 3.439\\ \hline
			$\rho = 0.2$ & 153 & 9.328& 3.441\\ \hline
			$\rho = 0.5$ & 162& 9.369 & 3.442\\ \hline 
		\end{tabular}
	\end{table}
\end{smaller}

\noindent\underline{\textbf{Effects of $\lambda$.}} Also, we investigate how $\lambda$ influences the insertion performance of~\modelName, where $\lambda$ is the pre-defined enlarging ratio for the fanouts of~\modelName's leaf node if a node adjustment is performed.
First, the~\modelName~is built with half of the dataset. Then, we vary the value of $\lambda$ and insert 100M keys to~\modelName. After that, search operations are conducted. 
Table~\ref{tab:lambda_effects} shows the average insertion time, the average results with different $\lambda$ values on the FB dataset. Results on other dtaasets are similar and thus omitted.
The insertion performance of~\modelName~is almost not influenced by $\lambda$. When the value of $\lambda$ is set to 2,~\modelName~achieves the best lookup time and the shortest tree structure.
\begin{smaller}
	\begin{table}[htb]
		\centering
		\smaller
		\caption{The effects of the hyperparameter $\lambda$}
		\label{tab:lambda_effects}
		\arrayrulecolor{black}
		\begin{tabular}{ l | r | r | r | r}
			\hline
			Param & \tabincell{c}{Insertion \\ time (ns)} & \tabincell{c}{lookup \\ time (ns)} & \tabincell{c}{Memory cost \\ ($10^9$ bytes)} & \tabincell{c}{Average\\height} \\ \hline
			$\lambda = 1.5$ & 483 & 182 & 8.924& 3.866\\ \hline
			$\lambda = 2$ & 487 & 180 & 8.911& 3.865 \\ \hline
			$\lambda = 4$ & 488 & 184 & 8.912 & 3.868\\ \hline
			$\lambda = 8$ & 485 & 185& 8.913 & 3.869\\ \hline 
		\end{tabular}
	\end{table}
\end{smaller}

\subsection{\modelName~vs RMI, RS and BU-Tree}
\label{subsec:exp_rmi_butree}

Our another experiment investigates why~\modelName~outperforms RMI, RS and the BU-Tree (Section~\ref{subsec:overall_idea}) on point queries. Note that~\modelName~and BU-Tree process point queries in the same two-step fashion: finding the relevant leaf node ($\mathtt{Step}$-$\mathtt{1}$), followed by search inside the leaf node ($\mathtt{Step}$-$\mathtt{2}$). Similarly, the search process with RMI and RS can be decomposed into two steps: the computation of the predicted position ($\mathtt{Step}$-$\mathtt{1}$) and the local search around the prediction ($\mathtt{Step}$-$\mathtt{2}$).
We create the BU-Tree using the whole dataset $\boldsymbol{P}$ with Algorithm~\ref{alg:build_bottom_up}.~\modelName~is built based on it. Then, we compare the breakdown time cost of the three models. Table~\ref{tab:exp_rmi_butree} reports the experimental results.

	\begin{table}[htb]
		\smaller
		\centering
		\caption{\modelName~vs RMI, RS \& BU-Tree}
		\label{tab:exp_rmi_butree}
		\begin{tabular}{ l | l | r | r | r }
			\hline
			Dataset & Model & $\mathtt{Step}$-$\mathtt{1}$ (ns) & $\mathtt{Step}$-$\mathtt{2}$ (ns) & Total (ns)\\ \hline
			
			\multirow{3}{*} {\textsf{FB}}
			& RMI & 139 & 76 & 215 \\ \cline{2-5}
			& RS & 264 & \textbf{41} & 305 \\ \cline{2-5}
			& BU-Tree & 386 & 210 & 596\\ \cline{2-5}
			& \modelName & \textbf{75} & 75 & \textbf{150} \\ \hline
			\multirow{3}{*} {\textsf{WikiTS}}
			& RMI & 138 & \textbf{37} & 175\\ \cline{2-5}
			& RS & 212 & 52 & 264  \\ \cline{2-5}
			& BU-Tree & 377 & 110 & 487 \\ \cline{2-5}
			& \modelName & \textbf{75} & 64 & \textbf{139} \\ \hline
			\multirow{3}{*} {\textsf{Logn}}
			& RMI & 135 & 73 & 208 \\ \cline{2-5}
			& RS & 101 & 31 & 132 \\ \cline{2-5}
			& BU-Tree & 273 & 57 & 330 \\ \cline{2-5}
			& \modelName & \textbf{73} & \textbf{43} & \textbf{116} \\ \hline
		\end{tabular}
	\end{table}

\noindent\underline{\textbf{\modelName~vs~RMI}}~\modelName~has clear advantage over RMI at $\mathtt{Step}$-$\mathtt{1}$. RMI is built through three linear stages or a cubic stage and two linear stages. Also, RMI needs to calculate the error bound when making predictions. 
In contrast, all~\modelName~versions have two layers of internal nodes and the step of locating the leaf node requires the calculations of two linear models only. 
Thus,~\modelName~requires less calculations.

At $\mathtt{Step}$-$\mathtt{2}$, RMI uses a local search to access the true position. The time cost of this step is correlated to the gap between the predicted position and the true position. In our experiments, the average gaps on the three datasets are 4.93, 2.97 and 53.9, respectively.~\modelName~needs to access one or more leaf nodes and do a calculations by the linear model per leaf node. Thus, the time cost of~\modelName's $\mathtt{Step}$-$\mathtt{2}$ depends on the number of traversed leaf nodes. According to our statistics, the average numbers of leaf nodes accessed per search on the three datasets are only 1.16, 1.05 and 1.02, respectively. Thus,~\modelName~is also more efficient at $\mathtt{Step}$-$\mathtt{2}$ in most cases.

\noindent \underline{\textbf{\modelName~vs~RS}} The average response times of the first step with RS are larger than the counterpart with~\modelName.
At the second step, RS additionally needs 41 ns, 52 ns, 31 ns on average to carry out the local search on the three datasets, respectively, whereas the response times of DILI at the second step are smaller or comparable.
Thus, RS's query performance is overall worse than DILI.

\noindent \underline{\textbf{\modelName~vs~BU-Tree}} Compared to the BU-Tree,~\modelName~incurs less time cost on both steps.
In particular at $\mathtt{Step}$-$\mathtt{2}$, the time gap between both trees is much larger. 
All in all,~\modelName~incurs considerably less total time as it is more efficient at finding the leaf nodes. This verifies the effectiveness of the design of~\modelName's internal nodes and the algorithm of building~\modelName~from the BU-Tree.

The BU-Tree's construction time is about 5 minutes. Based on that, building~\modelName~needs less than 1 minute extra. BU-Tree and~\modelName\\ 
consume 1.90-1.98$\times 10^9$ and 3.38-5.07$\times 10^9$ bytes respectively on the three datasets. Nevertheless,~\modelName~is much more efficient at search.

\subsection{Frequency and Effect of Node Adjustments}
\label{app:adjust_effect}

To investigate the effects of the adjusting strategy, we conduct experiments using~\modelName~and its variant~\ModelWOAD~which does not adopt the adjusting strategies when performing insertions. 
Following the Write-only workload evaluation process in Section 7.2 in the paper, both indexes are first built with half of the datasets. Then, for each dataset, we use both indexes to sequentially execute a Write-only workload containing 100M insertions and a Read-only workload containing 100M queries. The comparisons of their response times per insertion, memory costs, average heights, lookup time after insertions are shown in Table~\ref{tab:adjusting_effects} below. 
In particular, the third column of this table shows the average number of insertions per adjustment with~\modelName, which is the frequency of adjustments in DILI's leaf nodes.

\begin{smaller}
\begin{table}[htb]
	\centering
	\smaller
	\caption{The effects of the adjusting strategy}
	\label{tab:adjusting_effects}
	\setlength\tabcolsep{3pt}
	\arrayrulecolor{black}
	\begin{tabular}{ l | l | r | r | r | r  | r}
		\hline
		Dataset & Model & \tabincell{c}{Avg \# of \\ insertions per \\ adjustment} & \tabincell{c}{Insertion \\ time \\ (ns)} & \tabincell{c}{Memory \\ cost \\ ($10^9$ bytes)} &  \tabincell{c}{Avg\\ height} & \tabincell{c}{Lookup \\time \\ (ns)}\\ \hline
		
		\multirow{2}{*} {FB}
		& \ModelWOAD &  - & \textbf{442} & 9.31 & 3.91 & 187\\ \cline{2-7}
		& \modelName & 229.1 & 486 & \textbf{8.91}  & \textbf{3.86} & \textbf{180}\\ \hline
		\multirow{2}{*} {WikiTS}
		& \ModelWOAD & - & \textbf{349} & 5.86 & 3.58 & 172 \\ \cline{2-7}
		& \modelName & 758.4 & 361& \textbf{5.78} & \textbf{3.55} & \textbf{168}  \\ \hline
		\multirow{2}{*} {Logn}
		& \ModelWOAD & -& \textbf{304}& 3.71 & 3.11 & 130\\ \cline{2-7}
		& \modelName & 1304.6 & 310 & \textbf{3.67} & \textbf{3.09} & \textbf{126} \\ \hline
	\end{tabular}
\end{table}
\end{smaller}

Apparently, the adjusting strategy in \modelName's insertions will result in longer insertion time as it sometimes requires extra operations to collect all pairs covered by a node and train a new linear regression model. However, the adjusting strategy makes~\modelName~avoid more conflicts such that~\modelName~will have a shorter structure and achieves better lookup time and memory costs.

\subsection{A Possible Way to Reduce~\modelName's Construction Time}
\label{app:construction_time}

The most time-consuming step in~\modelName's construction is the greedy merging algorithm for getting the BU nodes at the bottom layer. Because we need to solve a MSE minimization problem for the union of each continuous two pieces, \emph{i.e.} $\mathcal{I}_u^k \bigcup \mathcal{I}^k_{u+1}$ and calculate the MSE to select the next two pieces to merge. Also, after merging $\mathcal{I}_u^k$ and $\mathcal{I}^k_{u+1}$ into a new piece, namely $\tilde{\mathcal{I}}_{u}^{k}$, we also need to conduct similar operations for $\mathcal{I}_{u-1}^{k} \bigcup \tilde{\mathcal{I}}_{u}^{k}$ (Algorithm 3 in the paper).

A direct yet effective approach to make the BU node generation more efficient is sampling. When a piece covers many keys, we could randomly or selectively sample part of the keys, \emph{e.g.,} select one key out of two, to get the linear regression model and calculate the cost. The sampling strategy makes little influence on the whole BU-tree node layout and the performance of the generated~\modelName. However, it will clearly reduce the construction time of the BU-tree and~\modelName. An experiment is conducted to show this.

We apply the sampling strategy on the BU-tree's construction as follows. When a piece $\mathcal{I}_u^k$ covers more than 8 keys, we only use half keys get the linear regression models. A~\modelName~is then built based on this BU-tree. As a comparison, we build another~\modelName~without adopting the sampling strategy. According to our statistics, the construction time of the former~\modelName~is 1 minute more less than that of the latter. Their comparison results on the lookup time are shown in Table~\ref{tab:sampling_effects}. Apparently, the lookup time of the~\modelName~with sampling is only slightly larger than that of the ordinary~\modelName.
\begin{table}[htb]
	\centering
	\smaller
	\caption{The lookup time of~\modelName~variants}
	\label{tab:sampling_effects}
	\setlength\tabcolsep{3pt}
	\arrayrulecolor{black}
	\begin{tabular}{ l | r | r | r | r | r}
		\hline
		Method& FB & WikiTS & OSM & Books & Logn\\ \hline
		\modelName-W-Sampling & 160 & 145 & 121 & 154 & 117\\ \hline
		\modelName & 150 & 139 & 117 & 148 & 116\\ \hline
	\end{tabular}
\end{table}

Considering the construction process of a~\modelName~will be executed once only, a slightly higher construction time is acceptable.

\subsection{Concurrent Insertions and Deletions in~\modelName}
\label{app:concurrent_insert}

\noindent It is possible that~\modelName~support concurrent data updates. It is noteworthy that either an insertion or deletion operation involves only one leaf node. The node adjustment of~\modelName~is much simpler than the rebalance operation of the B+Tree. Theoretically, the lock-free and lock-crabbing approaches can also be applied to~\modelName, in the same way as how they are applied to the B+Tree. For example, the lock-crabbing protocol for an insertion can be simply applied to~\modelName~as follows: 
\begin{enumerate}[label=(\Roman*)]
	\item Get the lock for the lowest leaf node, namely $\mathsf{N_D}$, covering the pair to be inserted.
	\item If $\mathsf{N_D}$ has an empty slot for the pair, put the pair here and release the lock for $\mathsf{N_D}$.
	\item If a conflict happens, the following steps are sequentially executed:
	\begin{itemize}[leftmargin=*]
		\item[1)] If a node adjustment is required, carry out the adjustment and exit. Otherwise, execute the following operations.
		\item[2)] Create a new leaf node $\mathsf{N'_D}$ and put the node at the conflicted slot.
		\item[3)] Get the lock for $\mathsf{N'_D}$.
		\item[4)] Release the lock for $\mathsf{N_D}$.
		\item[5)] Store the conflicting pairs in $\mathsf{N'_D}$.
		\item[6)] Release the lock for $\mathsf{N'_D}$.
	\end{itemize}
\end{enumerate}

The lock-crabbing protocal for a deletion is similar:
\begin{enumerate}[label=(\Roman*)]
	\item Get the locks for the lowest leaf node and its parent node, namely $\mathsf{N_D}$ and $\mathsf{N_P}$, respectively, covering the key to be deleted.
    
    \item Delete the corresponding pair from $\mathsf{N_D}$.
    
	\item If $\mathsf{N_T}$ is an internal node or $\mathsf{N_D}$ covers more than one pairs, release the locks for $\mathsf{N_D}$ and $\mathsf{N_P}$, and exit. Otherwise, execute the on the following operations:
        \begin{itemize}
            \item[1)] Suppose $\mathsf{N_P}.\boldsymbol{V}[k]$ stores the pointer to $\mathsf{N_D}$. Set $\mathsf{N_P}.\boldsymbol{V}[k] = p$.
            \item[2)] Delete the node $\mathsf{N_D}$.
            \item[3)] Release the lock for $\mathsf{N_P}$ and exit.
        \end{itemize}
\end{enumerate}
\fi

\end{document}